\documentclass[preprint2]{aastex63}

\usepackage{graphicx}
\usepackage{amsmath}
\usepackage[utf8]{inputenc} 
\usepackage[T1]{fontenc}    
\usepackage{url}            
\usepackage{booktabs}       
\usepackage{amsfonts}       
\usepackage{nicefrac}       
\usepackage{microtype}      
\usepackage{lipsum}
\usepackage{listings}
\usepackage{proghighlight}
\usepackage[frozencache=true, cachedir=.]{minted}
\usepackage{fancyvrb}

\setminted{breaklines=true}
\setminted{breakbytokenanywhere=true}

\newcommand{\taurex}{\texttt{TauREx}}
\newcommand{\trex}[1]{TauREx~#1}

\received{June 1, 2019}
\revised{January 10, 2019}
\accepted{\today}
\submitjournal{ApJ}

\shorttitle{TauREx 3}
\shortauthors{Al-Refaie et al.}
\graphicspath{{./}{figures/}}

\begin{document}

\title{TauREx 3: A fast, dynamic and extendable framework for retrievals}

\author{A. F. Al-Refaie}
\affiliation{Dept. Physics \& Astronomy, University College London, Gower Street, WC1E 6BT, London, UK}

\correspondingauthor{A. F. Al-Refaie}
\email{ahmed.al-refaie.12@ucl.ac.uk}

\author{Q. Changeat}
\affiliation{Dept. Physics \& Astronomy, University College London, Gower Street, WC1E 6BT, London, UK}

\author{I. P. Waldmann}
\affiliation{Dept. Physics \& Astronomy, University College London, Gower Street, WC1E 6BT, London, UK}

\author{G. Tinetti}
\affiliation{Dept. Physics \& Astronomy, University College London, Gower Street, WC1E 6BT, London, UK}

\begin{abstract}
\trex{3} is the next generation of the \taurex\ exoplanet atmospheric retrieval framework for Windows, Mac and Linux.
It is a complete rewrite with a full Python stack that makes it simple to use, high performance, dynamic and flexible. The new main \taurex\ program is built with modularity in mind, allowing the user to augment its functionalities with custom code and efficiently perform retrievals on their parameters.
We achieve this result by dynamic determination of fitting parameters, where \trex{3} can detect new parameters for retrieval from user code through a simple interface. \trex{3} can act as a library with a simple \pyth{import taurex} providing a rich set of classes and functions related to atmospheric modelling. A 10$\times$ speed-up in forward model computations is achieved compared to the previous version with a six-fold reduction in retrieval times while maintaining robust results. 
\trex{3} intends to act as a standalone, all in one package for retrievals while the \trex{3} Python library can build or augment a user's custom data pipeline easily.\\
\end{abstract}

\section{Introduction}

Characterisation of exoplanet atmospheres through spectroscopic methods has become a well established 
and rapidly growing field. Many retrieval codes now exist to solve the inverse forward model problem utilising varying methods such as the optimal estimation and Bayesian analysis with Markov Chain Monte Carlo \citep[MCMC][]{Jolliffe:2007vh} or nested sampling \citep{skilling2012}, a non-exhaustive list of such codes is NEMESIS: \cite{Irwin_nemesis}, CHIMERA: \cite{Line_chimera}, ARCiS: \cite{Ormel_arcis}, BART: \cite{harrington_bart}, petitRADTRANS: \cite{Mollire_petitrad}, Helios: \cite{Kitzmann, Lavie_helios}, POSEIDON: \cite{MacDonald_hd209}, Madhusudhan \& Seager: \cite{Madhusudhan_retrieval}, HyDRA: \cite{Gandhi_retrieval}, SCARLET: \cite{benneke_retrieval}, PLATON: \cite{Zhang_platon} and Pyrat-Bay: \cite{cubillos_pirat}).

Over the last few years, these methods have been successfully applied for a large number of cases. 
Initially designed for the analysis of hot-Jupiters \citep[e.g.][]{Tsiaras_pop_study_trans}, TauREx has been successfully applied to colder and smaller planets \citep[e.g.][]{tsiaras_55cnc,Tsiaras_k2-18,Edwards_LHS} as well as provide the basis for theoretical performance studies of current and future instruments \citep[e.g.][]{Venot_ariel_chem, Rocchetto_jwst, changeat2019complex, Katy_2020,Venot_2020_wasp43b, alfnoor}.

To date, the majority of atmospheric spectra of exoplanets have been observed with the Hubble Space Telescope (HST). The G102 and G141 grisms provide high signal to noise observations in the near infra-red covering two dominant water signatures \citep[e.g.][]{Kreidberg_2014, Tsiaras_pop_study_trans, Sing_2015,Swain_2009,Swain_2014, Line_hd209em, MacDonald_hd209, Madhusudhan_retrieval,Stevenson_2014,Stevenson_2017}. Observations from the Spitzer Space Telescope are sometimes combined to provide a more extensive wavelength coverage and some additional constraints on the carbon-bearing species: CH$_4$ at 3.6 $\mu$m and CO and CO$_2$ at 4.5 $\mu$m \citep[e.g.][]{Sing_2015, Stevenson_2017, Sheppard_2017, Evans_2017}. Combining the limited wavelength range covered by HST with Spitzer photometry,  allows for more precise temperature constraints in emission spectroscopy but comes at the risk of potentially introducing systematic errors \citep{Yip_lightcurve}. Additional observations from the ground can be used in atmospheric retrieval studies, providing better insight on the cloud properties, atomic lines and some of the metal oxides species \citep{MacDonald_hd209, Nikolov_2018, Sedaghati_2017}. Similarly, combining high-resolution ground-based spectroscopy with low-resolution space-based observations allows us to better leverage information contained in both during Bayesian retrievals \citep[e.g.][]{Brogi_2019}. For now, the focus of this article will deal with the retrieval of low-resolution spectroscopy only.

Recently, \trex{2} forward model and retrieval results have been verified against the NEMESIS and CHIMERA retrieval codes by \cite{barstow_comparison}. The study benchmarked retrievals from each code on mock forward models generated by the other codes. This cross-validation showed good agreement down to an observational noise floor of 30ppm.

As the field of exoplanet atmosphere sounding matures, the complexity of atmospheric forward models begins to outpace our ability to implement these models in existing `static' retrieval suites. Similarly, the growing complexity of state-of-the-art atmospheric forward models places a computational upper limit on what can realistically be included in a computationally intensive retrieval. This is particularly true for data obtained with the next generation of space telescopes (NASA/ESA - JWST  \citep{Gardner_JWSt,Bean_JWST} and the ESA/NASA - Ariel \cite{Tinetti_ariel} mission to be launched in 2021 and 2029 respectively. Here, forward models will have to evolve in order to cope with the new information content of these spectra. This puts more constraints on computing resources. In this context, it is necessary to develop the next generation of atmospheric retrieval frameworks able to cope effectively and efficiently with the increasing complexity of atmospheric modelling.

\trex{3} is a new atmospheric retrieval code for Windows, Mac and Linux written with a full Python 3 stack. It is a complete rewrite of \trex{2} and aims to improve upon its predecessor in three main areas: 1) performance in the computation of forward models, 2) flexibility in implementing 
and building of new forward models and 3) dynamic retrievals of any or all possible parameters.

We divide this paper in the following sections: \ref{sec:initial_setup}) Initial setup, where we explain the installation and basic run command; 
\ref{sec:framework}) The framework structure, discussing the architecture in more detail; 
\ref{sec:forward_models}) A description of the forward models; 
\ref{sec:opacity}) The available atmospheric opacities; 
\ref{sec:dynamicfit}) The Dynamic Parameters and retrieval setups; 
\ref{sec:instruments}) Instrument simulation modes; 
\ref{sec:performance}) Benchmarking the code against \trex{2};
\ref{sec:future}) Future Works and Discussions. \\
\\

\section{Initial Setup}
\label{sec:initial_setup}

Minimum requirements for \trex{3} is a Python 3 installation and \texttt{numpy}, all other dependencies required are automatically downloaded and installed. Full functionality (Equilibrium chemistry and
specific Mie scattering methods) will require additional FORTRAN and C++ compilers. To take advantage of \textit{Message Passing Interface} (MPI) nested samplers, an MPI library and compiler is required.

Installation of \trex{3} has been significantly simplified and requires only a single command:
\begin{minted}{bash}
$ pip install taurex
\end{minted}
Or if compiling from source:
\begin{minted}{bash}
$ cd TauREx3/ 
$ pip install .
\end{minted}

This gives the user access to a new program that can be run from anywhere in the operating system:
\begin{minted}{bash}
$ taurex --help
usage: taurex [-h] -i INPUT_FILE [-R] [-p] [-g] [-c] [-C] [--light]
              [--lighter] [-o OUTPUT_FILE] [-S SAVE_SPECTRUM]
\end{minted}

\noindent Running an input file \texttt{input.par}, storing and plotting a forward model can be done like so:
\begin{minted}{bash}
$ taurex -i input.par --plot -o myoutput.h5
\end{minted}

\noindent where `myoutput.h5' is an HDF5 file containing all generated data products such as spectra, contribution functions, molecular profiles etc. Performing a retrieval requires only the \texttt{retrieval} argument:
\begin{minted}{bash}
$ taurex -i input.par --plot -o myretrieval.h5 --retrieval
\end{minted}

\noindent The structure of the input file format has been reworked with each component of the atmosphere given an input header. For instance, the temperature profile defined under \texttt{[Temperature]} and chemical profile under \texttt{[Chemistry]}. These changes aim to improve readability significantly,  allowing the user to easily infer the type of atmosphere being computed and the nature of the retrieval conducted. Figure\,\ref{fig:t3_input} shows the input parameter file setup to perform retrievals for an example atmosphere with a free chemistry model\footnote{a heuristic chemical model where the mixing ratios of molecules are freely chosen} and an isothermal temperature profile. The modular nature of the input parameter file allows for easy addition and customization of parameters. 

\begin{figure}
    \centering
\begin{Verbatim}[frame=single,fontsize=\small]
[Global]
xsec_path=path/to/xsec/
cia_path=path/to/cia/

[Chemistry]
chemistry_type = free
fill_gases = H2,He
ratio = 0.17
    [[H2O]]
    gas_type = constant
    mix_ratio = 1e-3

[Temperature]
profile_type = isothermal
T=1300.0

[Pressure]
profile_type = simple
atm_max_pressure = 1e6
nlayers = 100

[Planet]
planet_type = simple
planet_mass = 1.0
planet_radius = 1.0

[Star]
star_type = blackbody
radius = 1.0

[Model]
model_type = transmission
    [[Absorption]]
    [[CIA]]
    cia_pairs=H2-H2,H2-He
    [[Rayleigh]]

[Observation]
observed_spectrum = hd209458b.txt

[Optimizer]
optimizer = nestle

[Fitting]
planet_radius:fit = True
planet_radius:bounds = 0.5,1.5
H2O:fit = True
H2O:bounds = 1e-12,1e-1
H2O:mode = log
\end{Verbatim}
    \caption{An example input file for \trex{3}}
    \label{fig:t3_input}
\end{figure}

One of the most powerful features within \trex{3} is its ability for the user
to extend the pipeline by injecting their own external atmospheric codes without modifying the main codebase. Extensions can be as simple as a new temperature profile to something more complex like a new radiative transfer model or an entirely new statistical sampling library for retrievals.
Within the input file, a user can point to their custom Python file and define key variables.
At run time \trex{3} will compile, determine new keywords, match them to the input, place them into the pipeline and retrieve any new fitting parameters the user has designed.
Taking a reasonable scenario, if a user created some new chemistry model for
\trex{3} in a separate \texttt{mychemistry.py} file, which has the form:
\begin{python}
class MyCustomChemistry(Chemistry):
   def __init__(self, param_one=10, 
                param_two='H2CO'):
      
    # Implement other features
    # and methods
   
\end{python}
We can now use this new chemistry model under the \texttt{[Chemistry]} header by pointing to the Python script and
\trex{3} will automatically create new input keywords based on the custom initialization keywords. An example of the now available input parameters is given below: 

\begin{Verbatim}[frame=single,fontsize=\small]
[Chemistry]
chemistry_type = custom
python_file = /path/to/mychemistry.py
param_one = 20
param_two = 'H2O'
\end{Verbatim}

This simple step fully integrates the custom model into the \taurex\ pipeline. A user can exploit automatic input keyword generation
to simplify the inclusion of external libraries such as statistical samplers or equilibrium chemistry models by defining the required arguments for a library as class initialization keywords.

In a sense, all of the atmospheric parameters, models and optimizers defined in this paper are simply the `batteries included'.

\section{Framework Structure}
\label{sec:framework}

\trex{3} provides flexibility and expand-ability by representing atmospheric parameters and contributions in the form of building blocks. These can be mixed and matched to form a complete forward model. 
The form of these building blocks is through abstract skeleton classes defined within \taurex. These classes (defined in Table \ref{tab:base-class}) provide a set of interfaces, in essence: a guarantee on what functions are provided, for other parts of the code to use. With this framework,
we can interconnect them knowing each object's responsibility.

\begin{table*}[]
\centering
\begin{tabular}{ll}
\hline\hline
Base class & Main-Purpose \\
\hline
\texttt{TemperatureProfile}  & Computes temperature profile \\

\texttt{Chemistry}           & Computes chemistry model    \\

\texttt{Gas}                 & Computes single species mixing ratio \\
                             &  for free-type chemistry model     \\

\texttt{PressureProfile}     & Computes pressure profile       \\

\texttt{Planet}              & Computes planet properties   \\

\texttt{Star}       & Computes stellar properties and flux \\

\texttt{Contribution}        & Performs a calculation on optical\\
                             & depth                            \\
\texttt{ForwardModel}        & Build and compute a forward model \\

\texttt{Spectrum}            & Provides some form of spectral data to fit against \\

\texttt{Optimizer}           & Performs retrievals \\

\texttt{Binner}              & Bins spectra to given grid \\

\texttt{Output}              & Handles file writes \\

\texttt{Instrument}          & Bins and generates noise \\

\hline
\end{tabular}
\caption{The base classes in \trex{3} \label{tab:base-class}}
\end{table*}

For example, when interpolating cross-sections we require temperatures for each layer of the atmosphere, an interpolator
does not require knowledge of the processes to build the profile, only that it \textit{can} be built and that a temperature profile is a result. This logic can (and is) applied to almost every aspect of the TauREx framework, from the chemistry and stellar profiles to the forward models, binning and optimizers used in retrievals. The framework makes very few assumptions about what is passed into the system, which increases the code's flexibility.

\begin{figure*}[t]
\centering
    \includegraphics[width=0.8\textwidth]{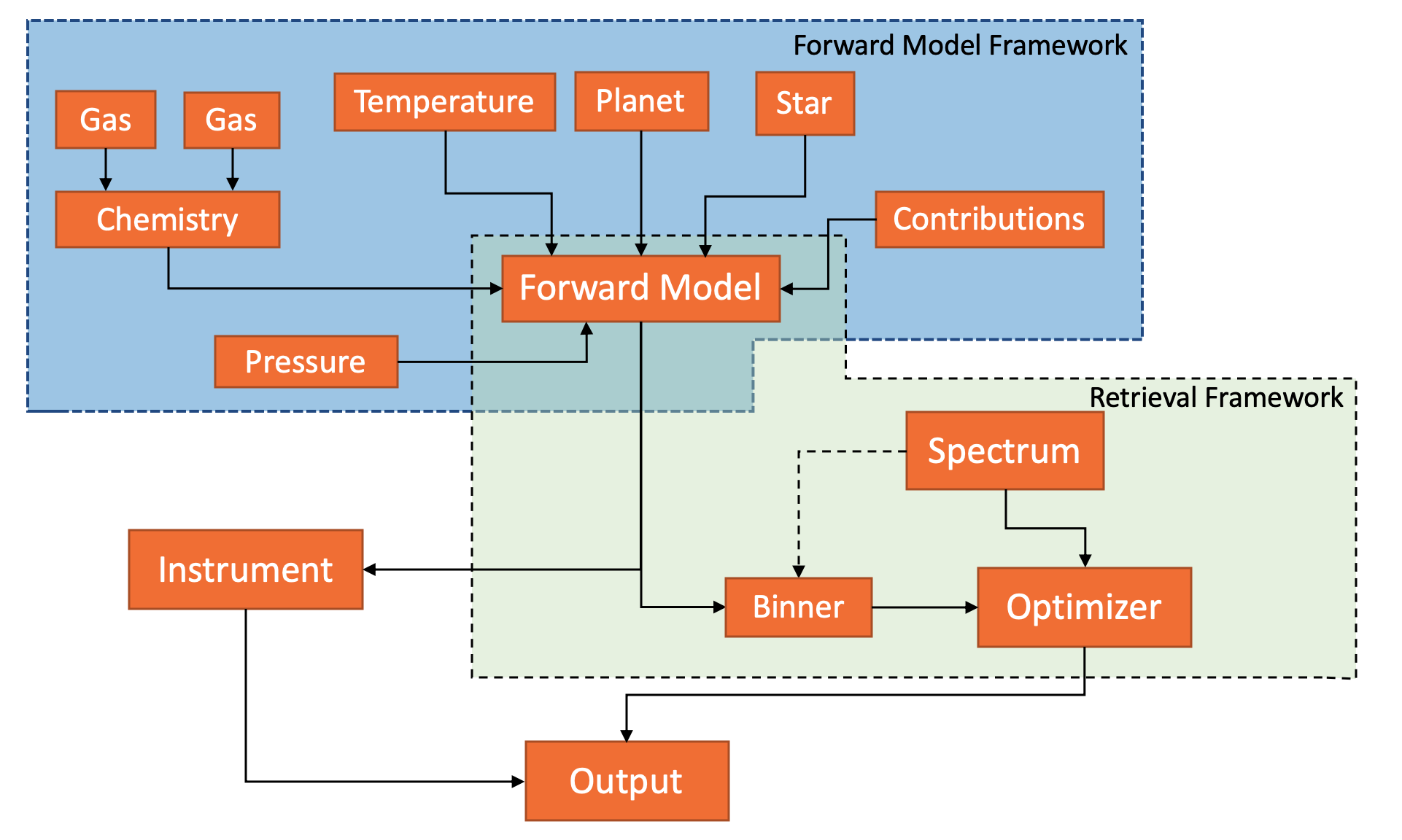}
\caption{The overall structure of \trex{3}. Highlighted is the two framework structure of the
complete framework. Each box describes a class from Table \ref{tab:base-class}, solid
arrows flowing out describe outputs, solid arrows flowing in describe inputs. Dashed arrows describe
creation of an object.}
\label{fig:taurexfm_framework}
\end{figure*}

Table \ref{tab:base-class} describes each of the available base classes within \trex{3}. Each of them
is created independently of the other. Figure \ref{fig:taurexfm_framework} describes the structure of \trex{3} and the interactions between different classes. \trex{3} consists of two separate frameworks:
The \textit{Forward Model Framework} and the \textit{Retrieval Framework}. The responsibility of the
Forward Model (FM) framework is to construct a \texttt{ForwardModel} class. The responsibility of the retrieval framework is to fit a \texttt{ForwardModel} against an observation (a \texttt{Spectrum}). The \texttt{ForwardModel} class
acts as a bridge between the two frameworks since at its most irreducible representation, it can be initialized, parameters can be read and set through a common interface (see Section \ref{sec:dynamicfit}), and a spectrum can be produced. The benefit of this is that the retrieval framework does not \textit{require} a forward model produced by the FM framework. A user can disregard it and pass in a custom \texttt{ForwardModel} class. The retrieval framework as a whole supports any arbitrary forward model. The only requirement is
that the forward model output and the observation to fit against match in their overall shape. Fitting for a single spectrum
will only function with a forward model that outputs a single spectrum. Fitting multiple spectra simultaneously requires
a forward model that returns multiple spectra. 
What the forward model computes internally is irrelevant to the optimizer.

\begin{figure*}[t]
\centering
    \includegraphics[width=0.8\textwidth]{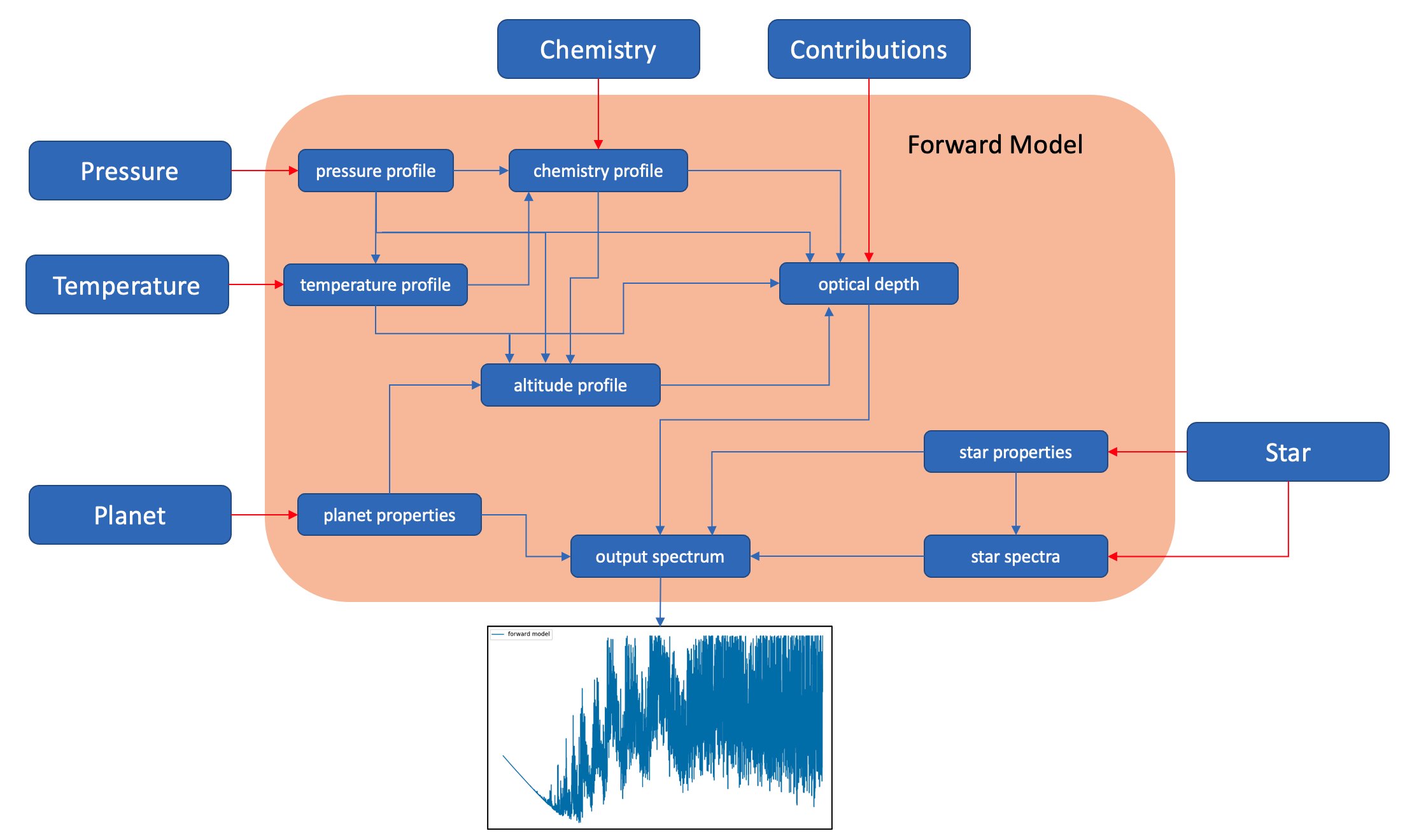}
\caption{A flow graph describing each atmospheric property class, their dependencies and the
resolution of these dependencies within the forward model. The large blue boxes describe a class from
Table~\ref{tab:base-class} and the red arrows describe the property they provide to the forward model.
Smaller grey boxes describe atmospheric properties produced within the forward model.
Arrow flowing inward display inputs required by particular property, arrows flowing out display output properties.}
\label{fig:taurexfm_flow}
\end{figure*}

With regards to the FM framework, its purpose is to build a forward model from replaceable atmospheric components dynamically. Some atmospheric properties have a natural dependency
on other aspects of the atmosphere. These dependencies occur through the interface methods for each class that provide
their own data. A chemistry profile may require temperature-pressure (TP) points,
a contribution function may require mixing profiles for each species. It is the responsibility of the \texttt{ForwardModel} class to aggregate all of these objects and interconnect them appropriately to resolve their dependencies. A basic implementation is provided by the higher-level abstract class \texttt{SimpleForwardModel}. Figure \ref{fig:taurexfm_flow} details how \texttt{SimpleForwardModel} moves data between each atmospheric parameter.

Generally, base and abstract classes do not assume the form of the coordinate system of the output data for each atmospheric component. The documentation for each of these base classes only describes what output data type is expected from each interface method, e.g. an array of values, a scalar, etc.
There are specific rules for concrete implementations. 
Between atmospheric properties, like-for-like dimensionality guarantees their ability to function with each other (e.g. any 1D TP profile can interconnect with any 1D chemistry profile). Mixing outputs from different coordinate
systems do not always guarantee compatibility. The currently implemented free chemistry profile only supports 1D TP profiles as an input and outputs 1D molecular profiles. The isothermal profile is coordinate agnostic and will output a 1D/2D or 3D profile based on the input pressure grid. Ultimately, the
forward model dictates how each component will function with each other. A 1D transmission
forward model will only work with components that output 1D data. 
The only concrete implementations available are 1D in this release. 
The FM framework standardizes all units to SI. It is up to the developer of new custom
classes to ensure that they convert inputs to their required units and convert outputs to the \trex{3} required SI units.

The consequence of this highly modular structure is that \trex{3} can also act as a library providing ready to use classes related to atmospheric modelling, such as cross-section interpolators, temperature profiles, chemistry models, contribution functions and optimizers to name a few. In essence, TauREx can be run as a standalone program requiring no additional coding or in a more interactive way via library imports for advanced and personalized usage. This is evident as once installed, 
\trex{3} can be imported as a library in any Python notebook, editor or shell:
\begin{minted}{python}
>>> import taurex
\end{minted}
\noindent Creating and computing a 2000\,K isothermal temperature profile is simply done:
\begin{minted}{python}
>>> from taurex.temperature import Isothermal
>>> iso = Isothermal(T=2000)
>>> iso.initialize_profile(nlayers=100)
>>> iso.profile
 array([1000., 1000., 1000., ..., 1000., 1000., 1000.])
\end{minted}
\noindent Loading and computing cross-sections for arbitrary temperatures and pressures is quickly done in three lines:
\begin{minted}{python}
>>> from taurex.opacity import PickleOpacity
>>> h2o = PickleOpacity('xsecpath/H2O.pickle')
>>> h2o.opacity(2000.0,1e1)  #2000 K 10 Pa
array([4.67879319e-29, 4.63121388e-28, 3.97618363e-28, ...,
       6.84398649e-34, 3.06316350e-34, 5.57666065e-34])
\end{minted}
\noindent These are just a small number of examples of the individual parts of \trex
that can be exploited for various purposes. The primary purpose is to combine
each of these parts to form our atmospheric model.
We can set up the rest of the planetary and stellar parameters for a Jupiter-like planet around a Sun-like star:
\begin{minted}{python}
>>> from taurex.planets import Planet
>>> from taurex.steller import BlackbodyStar
>>> planet = Planet(planet_mass=1.0, planet_radius=1.0)
>>> star = BlackbodyStar(temperature=5700, radius=1.0)
\end{minted}

\noindent generate a H$_2$, He atmosphere with constant H$_2$O:
\begin{minted}{python}
>>> from taurex.chemistry import TaurexChemistry
>>> from taurex.chemistry import ConstantGas
>>> chemistry = TaurexChemistry(fill_gases=['H2','He'])
>>> chemistry.addGas( ConstantGas('H2O',mix_ratio=1e-4) )
\end{minted}

\noindent build a transmission model with our atmosphere parameters:

\begin{minted}{python}
>>> from taurex.model import TransmissionModel
>>> tm = TranmissionModel(temperature=iso, chemistry=chemistry,
                          planet=planet,star=star,nlayers=100)
\end{minted}

\noindent add molecular absorption, collision-induced absorption (CIA) and Rayleigh scattering:
\begin{minted}{python}
>>> from taurex.contributions import *
>>> tm.add_contribution(AbsorptionContribution())
>>> tm.add_contribution(CIAContribution(cia_pairs=['H2-H2','H2-He']))
>>> tm.add_contribution(RayleighContribution())
\end{minted}
\noindent finally construct the model:
\begin{minted}{python}
>>> tm.build()
\end{minted}
\noindent and then run it:
\begin{minted}{python}
>>> from taurex.contributions import *
>>> wngrid, rprs, tau, extra = tm.model()
>>> rprs
array([0.01061007, 0.01073071, 0.01065356, ..., 0.01065933])
\end{minted}
\noindent Once built, our model can now be altered at will, such as
altering temperature and mix ratios:
\begin{minted}{python}
>>> tm['T'] = 1500.0
>>> tm['H2O'] = 1e-3
>>> wngrid, rprs, tau, extra = tm.model()
>>> rprs
array([0.01063282, 0.01074269, 0.01066373, ..., 0.01052198])
\end{minted}
\noindent This library works with more interactive flavours of Python such as IPython \citep{ipython} and Jupyter Notebook where forward models can be created, dynamically altered in real-time and used in retrievals.
Taking the library at a lower level, \trex{3} can be exploited simply for its dynamic fitting classes, allowing a developer to take advantage of the Bayesian retrieval framework for their codes and models.

\trex{3} also aims to conform to strict coding standards with full PEP-8 compliance. 
Full documentation is included with a suite of unit-tests used for debugging
and maintaining stability in the codebase during feature
development. Internally git-flow is used to manage contributions from multiple developers while maintaining compatibility.
Strict adherence to coding standards and source control is essential as often in the previous version; new features became isolated versions of the main code. With TauREx, we aim for continuous and compatible integration of new features into the main codebase. For external developers, we will use the fork-and-pull model.
Included is a developers guide which highlights the coding standards and rules for those wishing to contribute to the development and provides templates and examples.

\section{Forward Models}
\label{sec:forward_models}

All forward models derive from the abstract base class \texttt{ForwardModel}. 
This defines a simple skeleton, with an abstract \texttt{model} method 
that must return a native wavenumber grid, the result of the forward model, the
optical depth at each layer and any other extra information.

For this current release, only transmission and emission models are available but single and multi-scattering models will be included in future releases with additions to the framework to allow for more rapid implementation of these type of models from external libraries. 

One higher-level abstraction of the forward model is the \texttt{SimpleForwardModel} that
handles the majority of the setup for a 1D forward model by initializing and connecting each atmospheric component. It also computes the altitude profile $z$ at each layer $l$ using the expressions:
\begin{equation}
\begin{split}
z_l &= z_{l-1} + \Delta z_l \\
\Delta z_l &= -H_{l-1}log(\frac{P_l}{P_{l-1}}) \\
H_l &= \frac{k_B T_l}{\mu_l g_l} \\
z_0 &= 0 \\
\end{split}
\end{equation}
Where $P_l$, $T_l$, $\mu_l$, $g_l$ and $H_l$ are the pressure, temperature, mean molecular weight, acceleration due to gravity and scale-height at layer $l$ respectively. $\Delta z_l$ is the change in altitude from layer $l-1$ to $l$. $l=0$ is considered the bottom of the atmosphere and where \trex{3} defines the planetary radius.
The only method that must be defined
is the \texttt{path\_integral}. This design streamlined our transmission and emission implementation into a few lines of code.

For the application to exoplanet retrievals, we provide the basic forward models described from
previous versions \citep{Waldmann_taurex1,Waldmann_taurex2} for primary transits and secondary eclipses.

For the transit case, we model a 1D atmosphere where the altitude is parametrized by layers (default 100 layers). The total transit depth at wavelength $\lambda$ is given by:
\begin{equation}
    \Delta_{\lambda} = \frac{R_p^2 + a_{\lambda}}{R_s^2},
\end{equation}

\noindent where $R_p$ is the planet radius and $R_s$ is the parent star radius. We defined $a_{\lambda}$ as the wavelength dependant atmospheric depth with the form:
\begin{equation}
    a_{\lambda} = 2 \int_0^{z_{max}}(R_p+z)(1-e^{-\tau_{\lambda}(z)}) dz.
\end{equation}
where $z_{max}$ is defined as the altitude at the top of the atmosphere.
\noindent We define the wavelength dependant global optical depth $\tau_{\lambda}(z)$ as:

\begin{equation}
    \label{eq:tot_opt_dpth}
    \tau_{\lambda}(z) = \sum_i \tau_{\lambda,i}(z)
\end{equation}

\noindent where $\tau_{\lambda,i}$ denotes the optical depth for each absorber $i$, given by:

\begin{equation}
\label{eq:opt_dpth}
    \tau_{\lambda,i}(z) = \int_z^{z_{max}}\zeta_{i,\lambda}(z') \chi_{i}(z') \rho(z') dz',
\end{equation}

\noindent Here $\zeta_{m,\lambda}$ is the cross-section of a single absorbing species $i$, $\chi_{i}$ is the column density of the species $i$ and $\rho$, is the number density of the atmosphere. Opacity from collisional induced absorption (CIA) depends on a pair of molecular species; the integral instead has the form:

\begin{equation}
    \label{eq:cia_opt_dpth}
    \tau_{\lambda,i}(z) = \int_z^{z_{max}}\zeta_{i,\lambda}(z') \chi_{i}(z')\chi^{'}_{i}(z') \rho(z')^{2} dz',
\end{equation}
where $\chi_{i}$ and $\chi'_{i}$ are the column densities for their respective species in the pair $i$.
The emission model describes a plane-parallel atmosphere in which we integrate the emission from each layer to produce the final spectrum. The wavelength dependant intensity at the top of the atmosphere
from a viewing angle $\theta$:
    
\begin{equation}
    I(\tau = 0,\mu) = B_{\lambda}(T_s) e^{-\frac{\tau_s}{\mu}} + \int_0^{1}\int_0^{\tau_s}B_{\lambda}(T_{\tau}) e^{-\frac{\tau}{\mu}} d\tau d\mu,
\end{equation}

\noindent where we have defined $\mu=cos(\theta)$, $B_{\lambda}(T)$ as the Plank function at a given temperature T, with T$_s$ denoting the temperature at maximum atmospheric pressure and $\tau_s$ as the total optical depth from the planetary surface to the top of the atmosphere. We integrate for the cosine viewing angle $\mu$ using an $N$ point Gauss–Legendre quadrature scheme:
\begin{equation}
    I(\tau = 0) = \sum_i^{N} I(\tau=0,x_i)x_i w_i
\end{equation}
\noindent where $w_i$ and $x_i$ are our weights and abscissas respectively.
The final emission spectrum is expressed as:

\begin{equation}
    \frac{F_p}{F_s} = \frac{I(\tau = 0) }{I_s} \times \left(\frac{R_p}{R_s}\right)^2
\end{equation}

The user builds both transmission and emission models by providing a temperature profile, pressure parameters, chemistry model and contribution functions to the optical depth.

\subsection{Pressure Profiles}

Pressure profiles are represented as \texttt{PressureProfile} 
objects. Currently, only the  \texttt{SimplePressureProfile} class is
implemented, which for a given maximum pressure: $P_{max}$, minimum pressure: $P_{min}$ and number of layers $N_l$ computes an equally spaced logarithmic grid of $N_l+1$ pressures at each layer boundary. From this, we compute the central geometric pressure $P_l$ for each layer $l$ as:
\begin{equation}
    P_l = p_l\sqrt{\frac{p_{l+1}}{p_l}}
\end{equation}
\noindent where $p$ is the pressure at each layer boundary.

\subsection{Temperature Profiles}

\trex{3} adopts the layer-by-layer approach for the currently supported
temperature profiles. Included are a basic isothermal profile, 
a radiative two-stream approximation \citep{guillot}, a profile loaded from a file and a multi-point temperature profile. Their classes are given in Table \ref{tab:tpprofiles}.

\begin{table*}[t]
\centering
\begin{tabular}{ll}
\hline\hline
Class & Description \\
\hline
\texttt{TemperatureProfile}  & Base class \\

\texttt{Isothermal} & Isothermal temperature profile \\

\texttt{Guillot2010} & Radiative equilibrium \citep{guillot} \\

\texttt{Rodgers2000}   & Layer-by-layer     
\citep{rodger_retrievals} \\

\texttt{Npoint} & N-point temperature profile  \citep{Waldmann_taurex2} \\

\texttt{TemperatureFile} & Profile loaded from file \\
\hline

\end{tabular}
\caption{Available contributions in \trex{3} }\label{tab:tpprofiles}
\end{table*}

The previous version of \trex{2} included the 3-point and 4-point temperature profiles
that defined temperature points at different atmospheric pressures. Smoothing is applied
using a moving average kernel with user-definable window size.
These profiles are deprecated for the more general N-point profile which supports an arbitrary number of pressure and temperature points. Each point defined by the user dynamically generates new fitting parameters
for the retrieval. This aspect will be discussed further in Section \ref{sec:dynamicfit}.

\subsection{Chemistry}

\trex{3} supports equilibrium chemistry using the ACE FORTRAN chemistry code \citep{Agundez2012} using the thermochemical data by \citet{Venot_chem} and is installed if a suitable FORTRAN compiler is detected.
Here the C/O ratio and stellar metallicity can be retrieved.

For free chemistry models, \trex{3} can define different vertical mixing profiles for each molecule. For now, only three profiles are implemented, a constant mixing profile along the entirety of the atmosphere, a two-layer profile \citep{changeat2019complex} and a profile read from a file.
However, a custom profile can be used by implementing a \pyth{Gas} class and either adding it into the chemistry model through the \pyth{addGas} method or
defining the molecule with the \pyth{gas_type=custom} field and passing in the Python file.
All molecules included will generate their own set of fitting parameters (see Section \ref{sec:dynamicfit}), with each individual parameter for each molecular species capable of being retrieved.
The free chemistry model has been updated from \trex{2}, which only supported H$_2$, He atmospheres, to allow for more massive atmospheres with any number of molecules
through the \pyth{fill_gases} and \pyth{ratio} option. The
ratio term determines the portion of the remaining atmospheric volume relative to the first
fill molecule. We can, for example, arbitrarily define a heavy CO$_2$, CO, He atmosphere with a 4:3:1 ratio, 
a constant H$_2$O and two-layer CH$_4$ mixing profiles in the input file like so:

\begin{minted}{python}
[Chemistry]
chemistry_type = free
fill_gases = CO2,CO,He
ratio = 0.375,0.25

    [[H2O]]
    gas_type = constant
    mix_ratio=1e-4
    
    [[CH4]
    gas_type = twolayer
    mix_ratio_surface = 1e-6
    mix_ratio_top = 1e-8
    
\end{minted}

Custom chemistry models can also be used by setting \pyth{chemistry_type=custom} and pointing to an appropriate Python file. 
For all chemistry models, each molecule is considered either `active'
or `inactive'. This is automatically determined at runtime by the opacity caching system (discussed in Section \ref{sec:opacity}).
Discovery of absorption cross-sections for the particular molecule will label it as active; otherwise, they will be designated inactive. 
Active molecules will have a direct influence on the final spectrum, and molecular weight while inactive molecules will only affect the molecular weight.

\subsection{Contributions}

We can broadly generalize the optical depth $\tau$ from Equation \ref{eq:tot_opt_dpth} by considering it as a combination of multiple functions $C$:
\begin{equation}
    \label{eq:contrib}
    \tau(\lambda,z) = \sum_{i} C_{i}(\lambda,z)
\end{equation}
The most basic contribution function is pure absorption from Equation \ref{eq:opt_dpth}:
\begin{equation}
    C_i(\lambda,z) = \int \zeta_{i}(\lambda,z)w_{i}(z)\rho(z)dz
\end{equation}
\noindent where $\zeta_i$ is some cross-section $i$ weighted by some altitude dependant function $w_i$ and atmospheric density $\rho$. For molecular absorption, Mie scattering and Rayleigh scattering, $w_i$ is the column density $\chi_i$ of the absorbing species. 
We are not limited to this form though; we can redefine Equation \ref{eq:cia_opt_dpth} for collisionally induced absorption as:
\begin{equation}
    C_{i}(\lambda,z) = \int \zeta_{i}(\lambda,z)w_{i}(z)\rho(z)^{2}dz
\end{equation}
where $w_i$ is the product of column densities $w_i=\chi_i\chi'_i$.
\noindent We are free to use more exotic functions, for example, flat opacity clouds:
\begin{equation} \label{eq:clouds}
    C_{clouds}(\lambda,z) = 
    \begin{cases}
    \sigma       & \quad \text{if } P_{t} >= P(z) >= P_{b}\\
    0            & \quad \text{if } P(z) < P_{b} \\
    0            & \quad \text{if } P(z) > P_{t}
  \end{cases}
\end{equation}
\noindent where $\sigma$ is a user-defined opacity value in m$^{2}$, $P(z)$ is the pressure at altitude $z$, $P_t$, is the pressure at the top of the cloud deck and $P_b$ is the pressure at the bottom of the cloud deck. We note that the cloud model given by Equation \ref{eq:clouds} is discretized along $P(z)$. The choice of $P_t$ and $P_b$
will produce identical results when placed between two atmospheric layers. Discrete models are generally sufficient for currently available spectroscopic data, but next-generation telescopes will require
a comprehensive description of scattering from clouds. Future versions of \trex{3} will include more complex cloud models; however, a user is free to build on the current contribution framework to include these effects.

The calculation of the optical depth reduces to an array of these functions. We are free to dynamically create, add and remove them from the forward model according to our demands. Implementing new
scattering processes requires no modification of the underlying transmission and emission code.

These contribution functions are encapsulated in the \pyth{Contribution} class. 
The \pyth{prepare} method is called before each path integral and should perform initialization of any required data (e.g. load and interpolate cross-sections for molecular absorption). Our contribution function, $C$, is then implemented in the \pyth{contribute} method. 
A user can choose what type of contribution to add by inserting it
into a forward model using the \pyth{add_contribution} method or by defining it in the input file. Each contribution can have its own set of parameters that can be optimized during retrieval.
A list of available contributions is listed in Table\,\ref{tab:contributions}.

\begin{table*}[t]
\centering
\begin{tabular}{ll}
\hline\hline 
Class & Type of contribution \\
\hline
\texttt{Contribution}  & Base class\\

\texttt{AbsorptionContribution} & Molecular absorption  \\

\texttt{CIAContribution}        & Collisionally induced absorption (CIA) \\

\texttt{RayleighContribution}   & Rayleigh scattering       \\

\texttt{SimpleCloudsContribution}   & Grey clouds   \\

\texttt{BHMieContribution}   & Mie scattering \citep{bhmie}   \\

\texttt{LeeMieContribution}   & Mie scattering \citep{Lee_2013}   \\

\texttt{FlatMieContribution}   & Constant value opacity  \\
\hline
\end{tabular}
\caption{Available contributions in \trex{3} }\label{tab:contributions}
\end{table*}

Supplementing the pipeline with custom contributions in \trex{3} is only possible when used in the library form.
There is no custom option in the input file at the moment as conveying the option in the current
input file design would risk cluttering its structure.
A possible future option may be a user defined folder containing the user's set of custom Python files, which could then be automatically collected and parsed accordingly.

\subsection{Binning}

The \texttt{Binner} classes handle resampling spectra. \trex{3} provides the \texttt{FluxBinner} implementation, which bins both flux and uncertainties to a given grid with corresponding widths defined at class creation.
The implementation takes into account the relation between the spectral grid defined by the central bin: $\lambda_i$ and bin widths 
$\Delta\lambda_i$ and the resampling target grid defined by $\lambda_j$ and its corresponding bin-widths $\Delta\lambda_j$. For clarity we will define the minimum span of a spectral bin as: $\lambda^{-} = \lambda - \frac{\Delta\lambda}{2}$ and the maximum as $\lambda^{+} = \lambda + \frac{\Delta\lambda}{2}$.
For each spectral bin $\lambda_i$ that satisfies either of the rules:
\begin{equation}
\begin{split}
    \lambda^{-}_j &< \lambda^{-}_i < \lambda^{+}_j \\
    \lambda^{-}_j &< \lambda^{+}_i < \lambda^{+}_j \\
\end{split}
\end{equation}
\noindent we compute a corresponding weight $w_{ij}$ dependant on their occupancy within the resampling bin:
\begin{equation}
    w_{ij} = \frac{\min(\lambda_{j}^{+},\lambda_{i}^{+}) - \max(\lambda_{j}^{-},\lambda_{i}^{-})}{\Delta\lambda_i}
\end{equation}
\noindent where $\max$ and $\min$ are functions that select the largest and smallest value, respectively. For bins that fully lie within the resampling bin, this reduces to $w_{ij} = 1$.
The weighted mean of corresponding spectral fluxes $F_i$ is computed to produce the resampled flux $F_j$:
\begin{equation}
    F_j = \frac{\sum_iF_iw_{ij}}{\sum_iw_{ij}}
\end{equation}
If uncertainties $\sigma$ are included then the resampling takes the form of weighted propagation of uncertainties:
\begin{equation}
    \sigma_j = \sqrt{\frac{\sum_i w_{ij}^2 \sigma_i^{2}}{(\sum_iw_{ij})^{2}}}
\end{equation}
The algorithm can take into account overlapping bins and non-uniform grids. The class can be used outside of \trex{3} to bin spectra with and without uncertainties. Defining the binning in the input file will constrain the final spectrum to the given region. 

\section{Opacities}
\label{sec:opacity}
The previous version \trex{2} utilized both k-tables and absorption cross-sections, \trex{3} makes exclusive use of absorption cross-sections. The major optimizations within \trex{3} (discussed in Section \ref{sec:performance}) have meant that
k-tables perform no better computationally and are in-fact slower when taking multiple species into account. 

The absorption cross-sections come in the form of temperature-pressure-wavelength grids. \trex{3} includes two interpolation schemes. A faster
linear interpolation for temperature and pressure:

\begin{align} 
\sigma_{i}(T) &= \sigma_{i}(T_{1}) + m(T - T_{1})\text{,} \quad \\
&\text{where } \quad m = \frac{\sigma_{i}(T_{2}) - \sigma_{i}(T_{1})}{T_{2} - T_{1}}\text{.} \\
\sigma_{i}(P) &= \sigma_{i}(P_{1}) + m(P - P_{1})\text{,}\quad \\
&\text{where} \quad m = \frac{\sigma_{i}(P_{2}) - \sigma_{i}(P_{1})}{P_{2} - P_{1}}\text{.} 
\end{align}

\noindent where $\sigma_{i}$ is the absorption coefficient at wavelength $\lambda_i$, $P$ and $T$ are our chosen pressure
and temperature respectively and $P_1$, $P_2$, $T_1$ and $T_2$ are our pressure and temperature points on the grid
chosen so that $P_1 < P < P_2$ and $T_1 < T < T_2$. A second, more accurate, scheme \citep{HILL20131673} for temperature interpolation employs the form:
\begin{align*}\label{eq:exp-interp}
\sigma_{i}(T) &= a_{i}e^{- b_{i}/T}  \\
b_{i} &= \left( {\frac{1}{T_{2}} - \frac{1}{T_{1}}} \right)^{- 1} \ln{\frac{\sigma_{i}(T_{1})}{\sigma_{i}(T_{2})}} \\ 
\\
a_{i} &= \sigma_{i}(T_{1})e^{b_{i}/T_{1}}\text{.}
\end{align*}

Analysis by \cite{HILL20131673} and \cite{barton_h2o} have demonstrated interpolation residuals of less than 1.64\% for H$_2$O with a temperature sampling grid of $\Delta T=100$.
Testing on temperature grids with $\Delta T=200$ gives interpolation residuals of $\approx$ 3.4\% compared to the linear scheme of $\approx$ 11.2\%.

The \texttt{sympy} library\citep{sympy} was used to generate the most computationally efficient form of both interpolation schemes.
Either scheme can be activated using the \texttt{interpolation\_mode} keyword in the input file.
\begin{minted}{python}
[Global]
# Activate linear scheme
interpolation_mode = linear
# Activate exp scheme 
interpolation_mode = exp
\end{minted}

The exponential interpolation time is about 3$\times$ longer than the linear scheme due to the inclusion of the \texttt{exp} and \texttt{log} transcendental functions. The linear scheme is the default.
There is no standard or agreed-upon method for handling cross-section interpolation outside of its applicable temperature ranges. If the upper/lower  value lies outside of the cross-section's temperature or pre-calculated pressure range, we fix it to either the maximum or minimum temperature/pressure value. For the case when both, upper and lower, bounds are below the minimum applicable range, we return zero. When both pressure and temperature are above the maximum, the cross-section at maximum temperature and pressure is returned. We do not extrapolate wavelengths and all values outside the applicable wavelength range of the cross-section will return zero.

The wavenumber grid of the forward model is selected at runtime from whichever loaded opacity has the highest resolution. Every other opacity is then resampled to the chosen opacity's grid before any computations begin. This grid then becomes the `native' grid of the forward model.

\subsection{Formats}

\trex{3} supports the pickle format (based on Python object serialization)
used in \trex{2} for the absorption cross-sections but now also includes support
for the new HDF5 format from \cite{chubb2020exomolop}. The format comes with the option of streaming the coefficients used in the path integral directly from the HDF5 file, saving memory at the cost of approximately a 5-10$\times$ (dependant on the performance of the storage medium) degradation in performance from a significant increase in input/output (I/O) reads.
Reducing memory cost is advantageous when dealing with very high-resolution opacities. For example, opacities at $R=100,000$ covering a wavelength range of 0.3--15\micron\ with 20 temperature points and 20 pressure points requires roughly 2.6 GB of memory. For an 8GB workstation, this limits us to about 3--4 molecules. Streaming the opacities, offloads this memory requirement to local storage.
By default, all cross-sections are loaded into memory, but the streaming option can be activated using the \texttt{in\_memory=False} tag in the input file.
The \texttt{.dat} Exo-Transmit format \citep{exotransmit} are also supported. We recommend the ExoMol\footnote{\url{exomol.com}} project \citep{tennyson2012exomol, tennyson2016exomol} as a go-to source for cross-sections. It provides the molecular line lists calculated by ExoMol and/or provides links to the latest available line lists from third party sources. It further provides molecular broadening parameters and the ExoCross code \citep{exocross} used to build cross-sections for many molecular species in this study. The ExomolOP library \citep{chubb2020exomolop} provides pre-computed cross-sections in HDF5 format ready to used with \trex{3} and can be downloaded from the ExoMol website.

For collisionally-induced absorption, the HITRAN \citep{ RICHARD20121276,HITRAN2012,HITRAN2016, KARMAN2019160} \texttt{.cia} files are now supported and can be used directly rather than converting to pickle format. CIAs that contain different wavelength grids for different temperatures are also supported.

\subsection{Cache}

\trex{3} employs a lazy-loading scheme for absorption-coefficients facilitated by the \pyth{OpacityCache} class. This is a singleton
that is globally accessible to the entire program and will search a user defined folder and load absorption cross-sections when used. The cache requires a path
to be set through either the \texttt{xsec\_path} option in the input file or using the \pyth{set_opacity_path} method.
At this point a molecular cross-section object can be loaded into memory like so:
\begin{minted}{python}
>>> from taurex.cache import OpacityCache
>>> OpacityCache().set_opacity_path('path/to/xsec')
>>> h2o = Opacity()['H2O']
<taurex.opacity.pickleopacity.PickleOpacity at 0x106b54c50>
>>> h2o.opacity(temperature=2000.0,pressure=1e0)
array([8.73239546e-29, 2.57633453e-28, 8.40033984e-29, ...,
       6.84213835e-34, 3.05461786e-34, 5.57537933e-34])
\end{minted}
On initial run (depending whether streaming is used) accessing the
molecule using the square brackets operator can take a few seconds. Subsequent
calls will very quickly retrieve the cross-sections from cache:
\begin{minted}{python}
#First load of H2O cross-sections
>>> %timeit -r 1 -n 1 OpacityCache()['H2O']
1.24 s

#Second load of cross-sections
>>> %timeit OpacityCache()['H2O']
669 ns
\end{minted}
A user can expect the first run of the forward model to be delayed by up to a few seconds depending on the format and number of active molecules included.
When loading cross-sections from a path with multiple formats, loading priority is given to the HDF5 files before other formats are considered. 
The \pyth{CIACache} follows the same structure but for collisionally induced absorption files.

\section{Dynamic Parameters and Retrievals}
\label{sec:dynamicfit}

In the previous version of \trex{2}, fitting new physical parameters required explicitly hard-coding them into the retrieval. 
This approach is common to almost all retrieval codes that are currently available. There are significant limitations associated with this type of implementation. For instance, it does not scale well when
adding new parameters, as it significantly increases code complexity. This issue becomes much more apparent when attempting to merge features from multiple developers.
Apart from code complexity, another common issue is the discovery and
determination of `fittable' parameters. Taking chemistry as an example, equilibrium chemistry models simplify implementation  as they compute complex chemistry from a small and fixed number of parameters. However, when dealing with free chemistry models, the large number of free parameters becomes problematic. Furthermore, different molecules may have different mixing profiles, and we may wish to fit
different types of chemistry profiles, implying different `fittable' parameters and different prior configurations. In the previous version of \trex{2}, parameter specific priors were not supported, and all molecules
had to be fit to the same prior. At most, a later implementation of the two-layer model allowed the use of two different mixing profiles at the same time.

Finally, the scaling of the parameter space is commonly fixed in most codes. Generally, this is predetermined by the expected magnitude range of the parameter. Parameters such as trace gas volume mixing ratios have an extensive range of values and have their priors transform into logarithmic space. However, when it comes to the main constituent gases of the atmosphere (in the case of secondary atmospheres), it could be more appropriate to fit in linear space or to fit for ratios of components (such as H$_2$/He or H$_2$/N$_2$ ratios).
A choice in scaling often requires an explicit implementation for the specific parameter, which leads to more complexity in the codebase.

\trex{3} aims to solve this issue by dynamically determining the fitting parameters in a forward model.
Objects in \trex{3}, which have parameters to fit, inherit from the \texttt{Fittable} class. This includes \texttt{TemperatureProfile}, \texttt{Chemistry} and \texttt{ForwardModel} to name a few. The main purpose of the class is to discover, generate and advertise the
fitting parameters in the form of a Python dictionary with each item containing:

\begin{itemize}
    \item The name of the parameter
    \item The \LaTeX\ name of the parameter
    \item How it is read (its \texttt{fget})
    \item How it is written (its \texttt{fset})
    \item The default fitting space
    \item Whether to fit it by default
    \item Its prior bounds
\end{itemize}
The \texttt{fget} and \texttt{fset} functions are vital to this system as they provide the retrieval with a means to sample parameters without knowledge of what the parameters are and from where they originated.
The last three items in the list above are used by the optimizer to control the nature of the prior transform for the parameter and can be altered at runtime.

Taking the \texttt{Guillot2010} temperature profile as an example, we can determine the fitable parameters by querying the object:
\begin{minted}{python}
>>> guillot = Guillot2010(T_irr=1200)                                   
>>> params = guillot.fitting_parameters()
>>> params.keys()
dict_keys(['T_irr', 'kappa_irr', 'kappa_v1', 'kappa_v2', 'alpha'])
\end{minted}

\noindent We can read the parameter name and latex form:
\begin{minted}{python}
>>> params['T_irr'][0]
T_irr
>>> params['T_irr'][1]
'$T_\\mathrm{irr}$'
\end{minted}

\noindent and we can show that these getters and setters have a direct influence on the temperature profile:
\begin{minted}{python}
>>> guillot.equilTemperature
1200.0
>>> params['T_irr'][2]()
1200.0
>>> params['T_irr'][3](1300.0)
>>> params['T_irr'][2]()
1300.0
>>> guillot.equilTemperature
1300.0
\end{minted}

\noindent and finally we can obtain the default fit space, whether it is enabled and what its default prior bounds are:
\begin{minted}{python}
>>> params['T_irr'][4]
linear
>>> params['T_irr'][5]
True
>>> params['T_irr'][6]
[1300, 2500]
\end{minted}

This approach gives the retrieval scheme all the necessary information required to sample without explicitly defining parameters in the code.
However, this form is relatively cumbersome to use. When placed inside a \texttt{ForwardModel} class, the parameters are collected into a unified parameter pool allowing all of them to be easily accessed:
\begin{minted}{python}
>>> tm = TransmissionModel(temperature_profile=guillot)
>>> tm.build()
>>> tm.fittingParameters.keys()                                                                                         
dict_keys(['planet_mass', 'planet_radius', 'planet_distance', 
'atm_min_pressure', 'atm_max_pressure', 'T_irr', 'kappa_irr', 
'kappa_v1', 'kappa_v2', 'alpha', 'H2O', 'CH4', 'He_H2'])
\end{minted}

We see that our \texttt{Guillot2010} profile has been detected by the forward model and been added to the pool. The same occurs if we use \texttt{Isothermal} instead:

\begin{minted}{python}
>>> tm_iso = TransmissionModel(temperature_profile=Isothermal(T=1000))
>>> tm_iso.build()
>>> tm_iso.fittingParameters.keys()                                                                                     
dict_keys(['planet_mass', 'planet_radius', 'planet_distance', 
'atm_min_pressure', 'atm_max_pressure','T', 'H2O', 
'CH4', 'He_H2'])
\end{minted}

The other parameters arise from the default profiles used when nothing else is defined in the forward model. The \texttt{ForwardModel} classes  provide a very simple method of accessing these parameters with the square bracket operator:

\begin{minted}{python}
>>> tm['T_irr']
1300.0
>>> tm['T_irr'] = 1400.0
>>> tm['T_irr']
1400.0
>>> guillot.equilTemperature
1400.0
\end{minted}

The process of creating fitting parameters is simple. The \texttt{Fittable} class provides two ways of defining them. 
The first method is provided by the \pyth{@fitparam} decorator. This decorator behaves identically to the 
    builtin Python \pyth{property} with extra arguments given in Table \ref{tab:fit-params}

\begin{table*}
\centering
\begin{tabular}{lrr}

\hline \hline
Argument & Description & Values\\
\hline
\texttt{param\_name} & Parameter key name & \\
\texttt{param\_latex} & \LaTeX\ name & \\
\texttt{default\_mode} & Default fitting space & either `linear' or `log'\\
\texttt{default\_fit} & Retrieve by default & \texttt{True} or \texttt{False}\\
\texttt{default\_bounds} & Default prior bounds & [bound min, bound max]\\
\hline
\end{tabular}
\caption{Arguments for the \texttt{@fitparam} decorator}\label{tab:fit-params}
\end{table*}

We can create a custom temperature profile \pyth{FoobarProfile} with parameter \texttt{Foobar\_T} as seen in Figure \ref{lst:custom-temp-profile}:
\begin{figure}
    \begin{minted}{python}
    class FoobarProfile(TemperatureProfile):
    
        def __init__(self,foobar = 1000):
            super().__init__()
            
            self._foobar = foobar
        
        # Other code.....
        
        @fitparam(param_name='Foobar_T',
                  default_mode = 'linear',
                  default_bounds=[1200.0,1500.0])
        def myFooBar(self):
            return self._foobar
        
        @myFoobar.setter
        def myFooBar(self,value):
            self._foobar = value
        
        # More code.....
        
    \end{minted}

    \caption{Our custom temperature profile}
    \label{lst:custom-temp-profile}
\end{figure}

We can get and set the parameter like a normal Python \pyth{property}:
\begin{minted}{python}
>>> foo = FoobarProfile()
>>> foo.myFooBar
1000.0
\end{minted}

\noindent and by adding it to a forward model we demonstrate that it is detected:

\begin{minted}{python}
>>> tm = TransmissionModel(temperature_profile=foo)
>>> tm.fittingParameters.keys()                                                                              
dict_keys(['planet_mass', 'planet_radius', 'planet_distance', 
'atm_min_pressure', 'atm_max_pressure','Foobar_T', 'H2O', 
'CH4', 'He_H2'])
>>> tm['Foobar_T']
1000.0
>>> tm['Foobar_T'] = 1200.0
>>> foo.myFoobar
1200.0
\end{minted}
\noindent Now, our custom class is ready for retrievals.

The second method is with the \pyth{add_fittable_param} method. This has similar arguments
to the \pyth{@fitparam} decorator but must be explicitly provided with the getter and setter functions. This method allows for dynamic fitting parameters that alter depending on how it was created (such as profiles with arrays).
\pyth{NPoint} demonstrates this dynamic nature, where the fitting parameters change depending on
the number of temperature and pressure points set:
\begin{minted}{python}
# Two point profile
>>> twop = NPoint()
>>> twop.fitting_parameters().keys()
dict_keys(['T_surface', 'T_top', 'P_surface', 'P_top'])
# Four point profile
>>> fourp = NPoint(temperature_points=[1000.0,2000.0], 
                    pressure_points=[1e2,1e0])
>>> fourp.fitting_parameters().keys() 
dict_keys(['T_surface', 'T_top', 'P_surface', 'P_top', 'P_point1', 'P_point2', 'T_point1', 'T_point2'])
\end{minted}

\subsection{Optimization}

For the retrievals, \trex{3} comes with the methods given in Table\,\ref{tab:retrieval-meth} built-in. Nestle \citep{nestle} is a pure Python Bayesian nested sampler that is automatically downloaded and installed with \trex{3} and can immediately be used for retrievals. Multinest \citep{multinest}, Polychord \citep{polychord} and dyPolychord \citep{dypolychord1,dypolychord2} require their respective FORTRAN libraries to be built and Python wrappers to be installed before they can be used in \trex{3}. They will automatically be detected at runtime once they are installed. A user can include their sampler through the \pyth{optimizer = custom} flag in the input file. 
\begin{table*}[]
\centering
\begin{tabular}{lll}
\hline\hline
Method & \trex{3} class & Ref \\
\hline
MultiNest & \texttt{MultiNestOptimizer}  & \cite{multinest} \\

Nestle & \texttt{NestleOptimizer}     & \cite{nestle}    \\

Polychord & \texttt{PolychordOptimizer}  & \cite{polychord} \\

dyPolychord & \texttt{dyPolychordOptimizer}  & \cite{dypolychord1,dypolychord2}       \\

\hline
\end{tabular}
\caption{Supported retrieval libraries in \trex{3} }\label{tab:retrieval-meth}
\end{table*}

The base \pyth{Optimizer} class is responsible for collecting the fitting
parameters from the forward model, updating the model using the fitting parameters from the sampler and computing the likelihood. The responsibility of the optimization is handled by the \pyth{compute_fit} method, which must be defined in concrete classes.
The user can inform the optimizer which fitting parameter to retrieve though its
\pyth{enable_fit} and \pyth{disable_fit} methods.
The optimizer also handles the parameter space conversion by wrapping the \pyth{fget}
and \pyth{fset} with an appropriate conversion function, determined by the
\pyth{default_mode} attribute within each fitting parameter. This conversion can be altered by
the user programmatically using the \pyth{set_mode} method. Currently, only linear space and log-10 space is supported.
Fits can also be defined in the input file under the \pyth{[Fitting]} section. The input file is dynamic and is capable of setting user-defined fitting parameters as well. Take our profile from Figure \ref{lst:custom-temp-profile}  as an example, we can fit for \texttt{Foobar\_T} in log scale with bounds 200.0, 1000.0 like so:

\begin{minted}{python}
[Temperature]
profile_type = custom
python_file = foobar.py
foobar = 1500.0

[Fitting]
Foobar_T:fit = True
Foobar_T:mode = log
Foobar_T:bounds = 200.0, 1000.0
\end{minted}

It should be noted that the bounds are always in linear space. The log conversion
is automatically handled by the optimizer. Currently, only uniform priors are supported.

\subsection{Model Rejection}

During sampling, there may be regions within the parameter space that are non-physical. Non-physical forward models can include atmospheres that have
greater than unity mixing ratios or multi-point temperature profiles that have inverted pressure points. Sampling is wasted on these regions as they may create accidental modes in the solution if they inadvertently produce ``correct" spectra. To combat this, \trex{3} provides the \pyth{InvalidModelException}
exception. During retrieval, any profile or chemistry model can trigger this exception, forcing the log-likelihood to the lowest possible value. For both nested
sampling and classical MCMC sampling, this results in an overall avoidance of these regions, which should result in slightly faster sampling.

\section{Instruments}
\label{sec:instruments}
\begin{center}
\begin{figure}[]
\centering
    \includegraphics[width=0.45\textwidth]{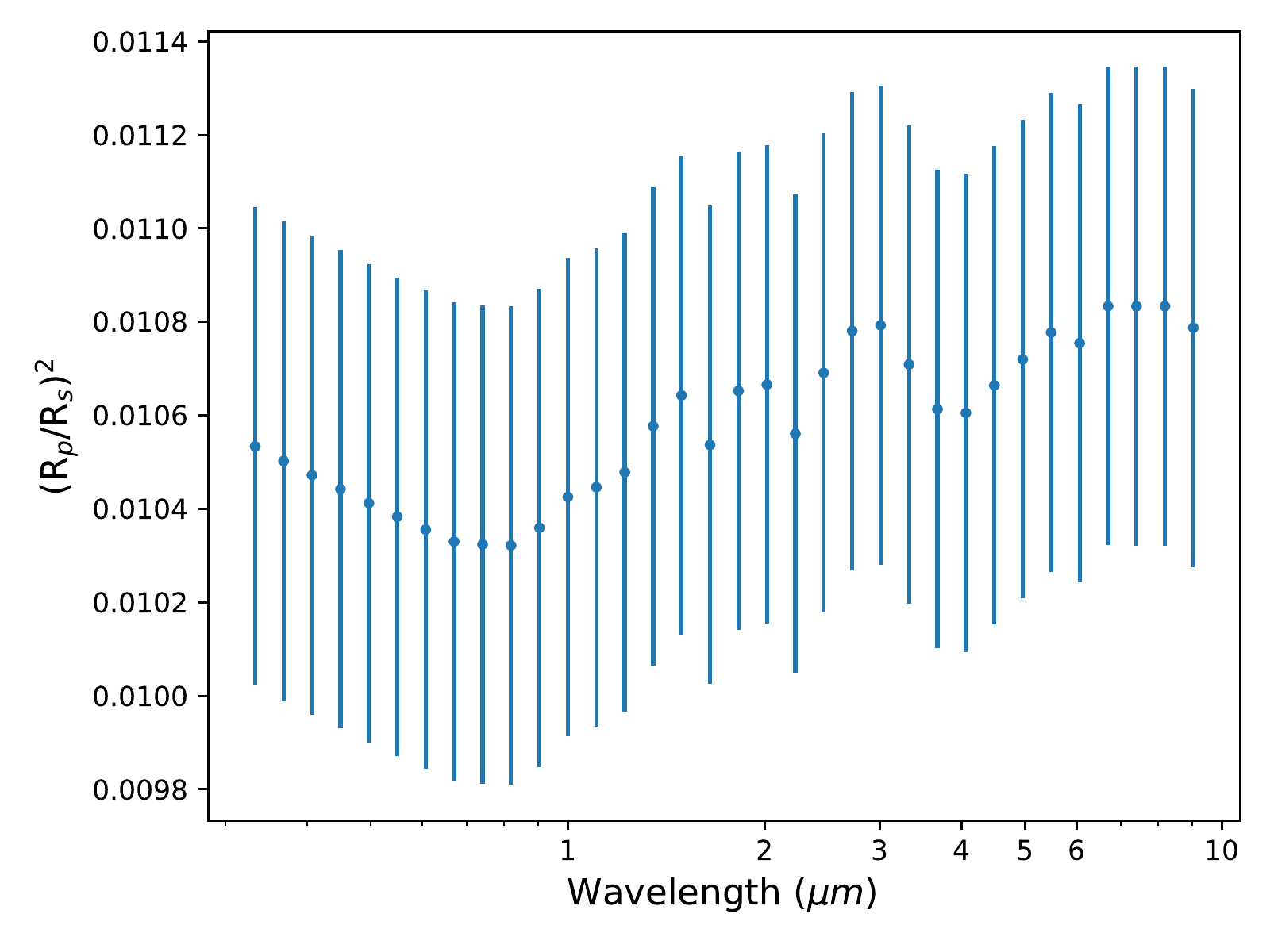}
    \includegraphics[width=0.45\textwidth]{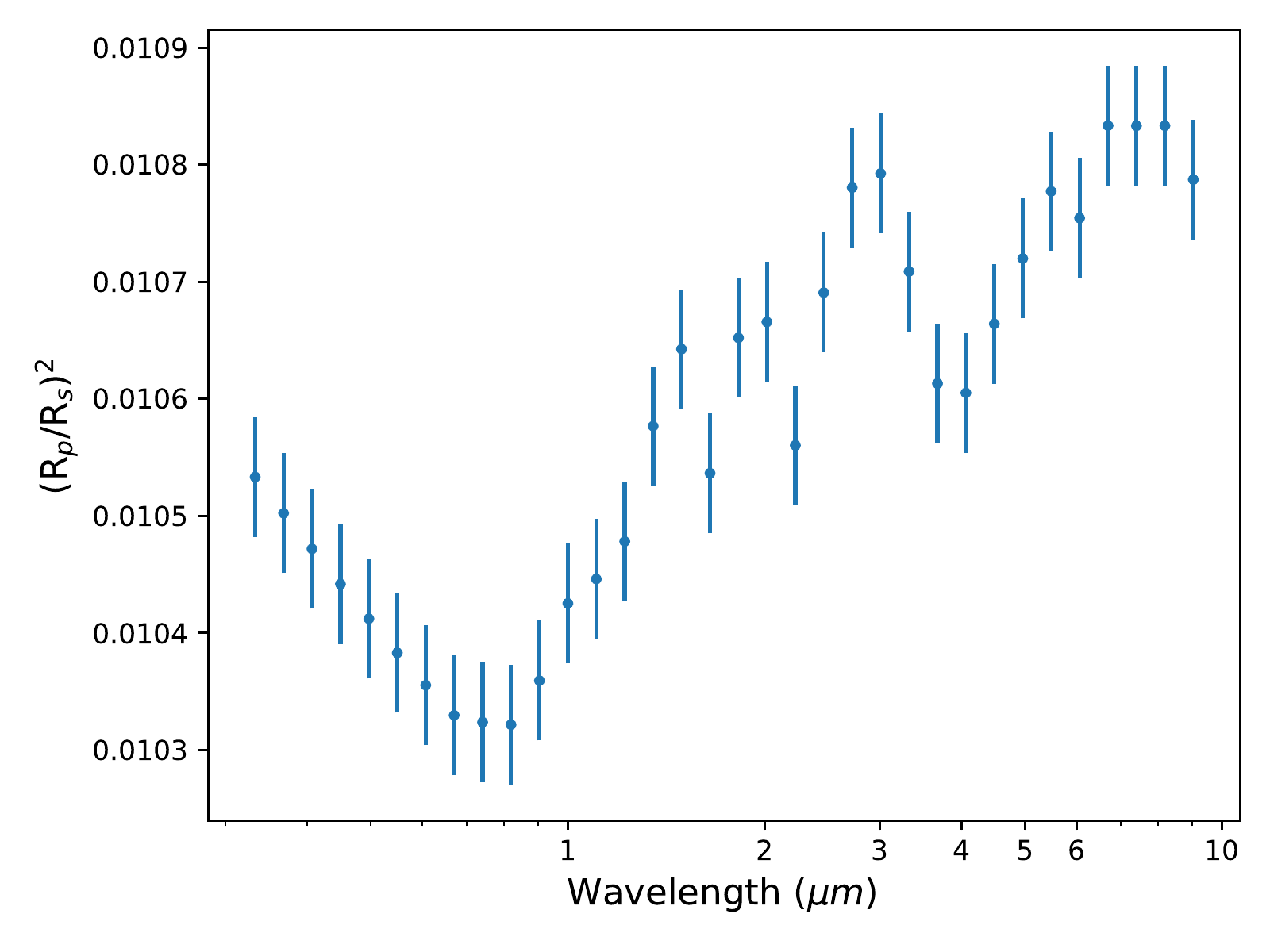}
    \includegraphics[width=0.45\textwidth]{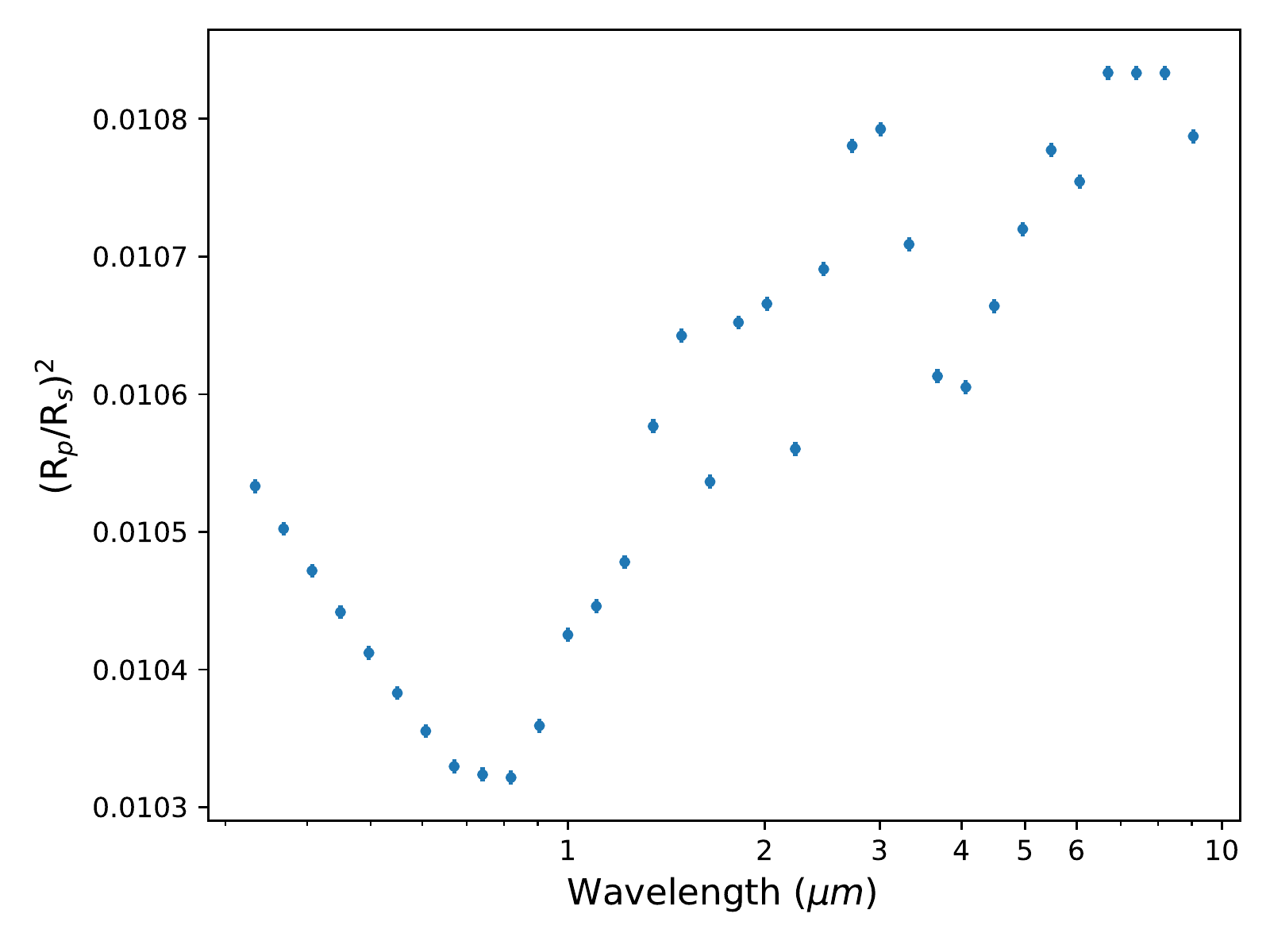}
\caption{Plots of same forward model applied with the signal-to-noise systematics model binned to $R=10$ for visual clarity. Points denote the spectrum and error bars denote the noise. The forward model is a clear isothermal atmosphere with $T=2000$, $0.001$\% H$_2$O (\cite{barton_h2o} and \cite{polyansky_h2o}) and Rayleigh scattering \citep{cox_allen_rayleigh}.
Each plot shows an increasing value of signal-to-noise ($SNR$) with Top: $SNR=1$, Middle: $SNR=10$
and Bottom: $SNR=100$}
\label{fig:snr}
\end{figure}
\end{center}

\begin{center}
\begin{figure}[]
\centering

    \includegraphics[width=0.45\textwidth]{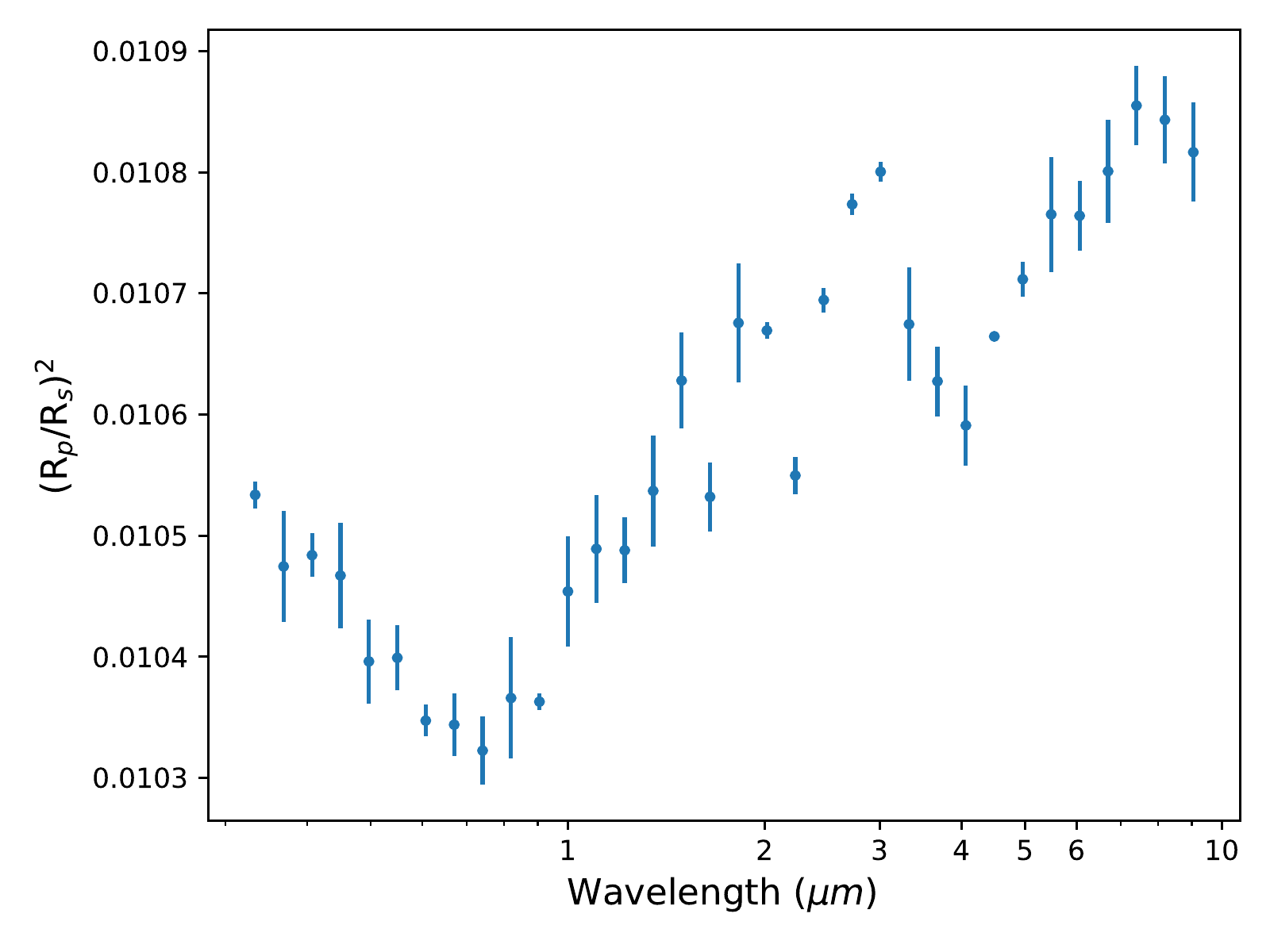}
\caption{A custom instrument applied to the same forward model from Figure \ref{fig:snr}. This instrument model generates random Gaussian noise and scatters the spectra.  }
\label{fig:snr_rand}
\end{figure}
\end{center}

One of the new features within the \taurex\ pipeline is the instrument model
simulator. If defined, it passes the result of a forward model simulation
into an instrument noise model and generates a new binned spectrum with instrument noise and systematics. The number of observations can also be passed in to simulate further the effect of stacking multiple observations.
Currently, a generic signal-to-noise ratio (SNR)  model is included which computes normally distributed noise $\mathcal{N}$ at all wavelengths for a given $SNR$ based on the simple relation:
\begin{equation}
    \mathcal{N} = \frac{S}{SNR} 
\end{equation}
\noindent with $S$ representing the maximum value (generally the maximum transit/eclipse depth) in the spectrum. The final noise $\mathcal{N}'$
is computed based on the number of observations $n$ passed in by the user:
\begin{equation}
    \mathcal{N}'=\frac{1}{\sqrt{n}}\mathcal{N}
\end{equation}
Figure \ref{fig:snr} shows the same forward model binned to $R=50$ and applied with increasing values of $SNR$. The forward model is arbitrary, but the same for all of the plots. This instrument model produces only symmetric uncertainties.
The user can provide custom noise models for instruments by passing in
the \pyth{instrument=custom} flag as well as pointing to the correct Python file. An example is provided in Figure \ref{fig:snr_rand}, where a custom instrument model is built that generates random Gaussian noise and then subsequently scatters the spectrum according to its uncertainties.

The instrument model in the \taurex\ pipeline can be used to bypass loading in an observation and to perform, instead, a retrieval directly on the simulated observation.  These simulated observation retrievals provide the user with a convenient toolbox to estimate the retrievability of atmospheric parameters giving a range of telescope/instrument setups (e.g. optimizing future JWST/ARIEL observations).


\section{Benchmark}
\label{sec:performance}

\subsection{Computational}
In the previous version of \trex{2}, the framework was written using a combination of Python for the general codebase and C++ for the heavy computational work. \trex{3} has switched to a full Python stack, which includes the computation of the path integral. Common knowledge dictates that Python is slower than
compiled languages such as C++ and FORTRAN. However, 
we mitigate slower computational speeds using a suite of available libraries to speed up and even match performance compared to compiled languages, without sacrificing the flexibility of Python.

\trex{3} fully leverages \texttt{numpy}~\citep{numpy} for its array vectorization capabilities and \texttt{numexpr}~\citep{numexpr} for faster NumPy operations. The brunt of the calculation exploits the \texttt{numba}~\citep{numba} library, which just-in-time (JIT) compiles the path integral code for even faster performance. An additional performance gain is achieved in the opacities calculation: once each opacity has been interpolated and weighted, they are fused into a single cross-section from which the path integral can then be calculated. This method significantly improves the scaling of performance with the number of molecules. This optimization is also applied to Rayleigh Scattering and CIA.

Multi-threading was not used in \trex{3} as it does not benefit retrievals. It is generally better to let an MPI sampler (such as Multinest) have more cores to sample the forward models in parallel as one can generally get linear scaling with core counts. For both Multinest and PolyChord, this holds as long as the sampling is the dominant computational bottleneck.

For our forward model benchmarks, we test on a MacBook Pro 2018 with a 2.3 GHz Intel Core i5. The absorption opacities were computed from the Exomol line lists \citep{tennyson2012exomol,tennyson2016exomol} using the Exocross \citep{exocross} FORTRAN code at a wavelength range of 0.3--15\micron\ with
resolution of $R=10000$. The k-tables were also generated from the same line lists at $R=100$ using 20 Gaussian quadrature points. Linear interpolation is used in all benchmarks.
The particular sources for the absorption line-lists and CIA opacities are listed in Table \ref{tab:opacity-source}. The timings for the forward model were conducted using the \pyth{timeit} module.

\begin{table*}[]
\centering
\begin{tabular}{lll}
\hline\hline
Opacities & Type &  Ref.  \\
\hline

H$_2$-H$_2$  & CIA & \cite{abel_h2-h2}, \cite{fletcher_h2-h2} \\
H$_2$-He   & CIA & \cite{abel_h2-he} \\
H$_2$O     & Abs. & \cite{barton_h2o}, \cite{polyansky_h2o} \\
CH$_4$     & Abs. & \cite{hill_xsec}, \cite{exomol_ch4} \\
CO       & Abs. & \cite{li_co_2015} \\
CO$_2$     & Abs. & \cite{rothman_hitremp_2010} \\
NH$_3$     & Abs. & \cite{Yurchenko_nh3} \\
\hline
\end{tabular}
\caption{List of opacities used for our benchmarks}\label{tab:opacity-source}
\end{table*}

\begin{figure*}[p]
\centering
    \includegraphics[width=0.98\textwidth]{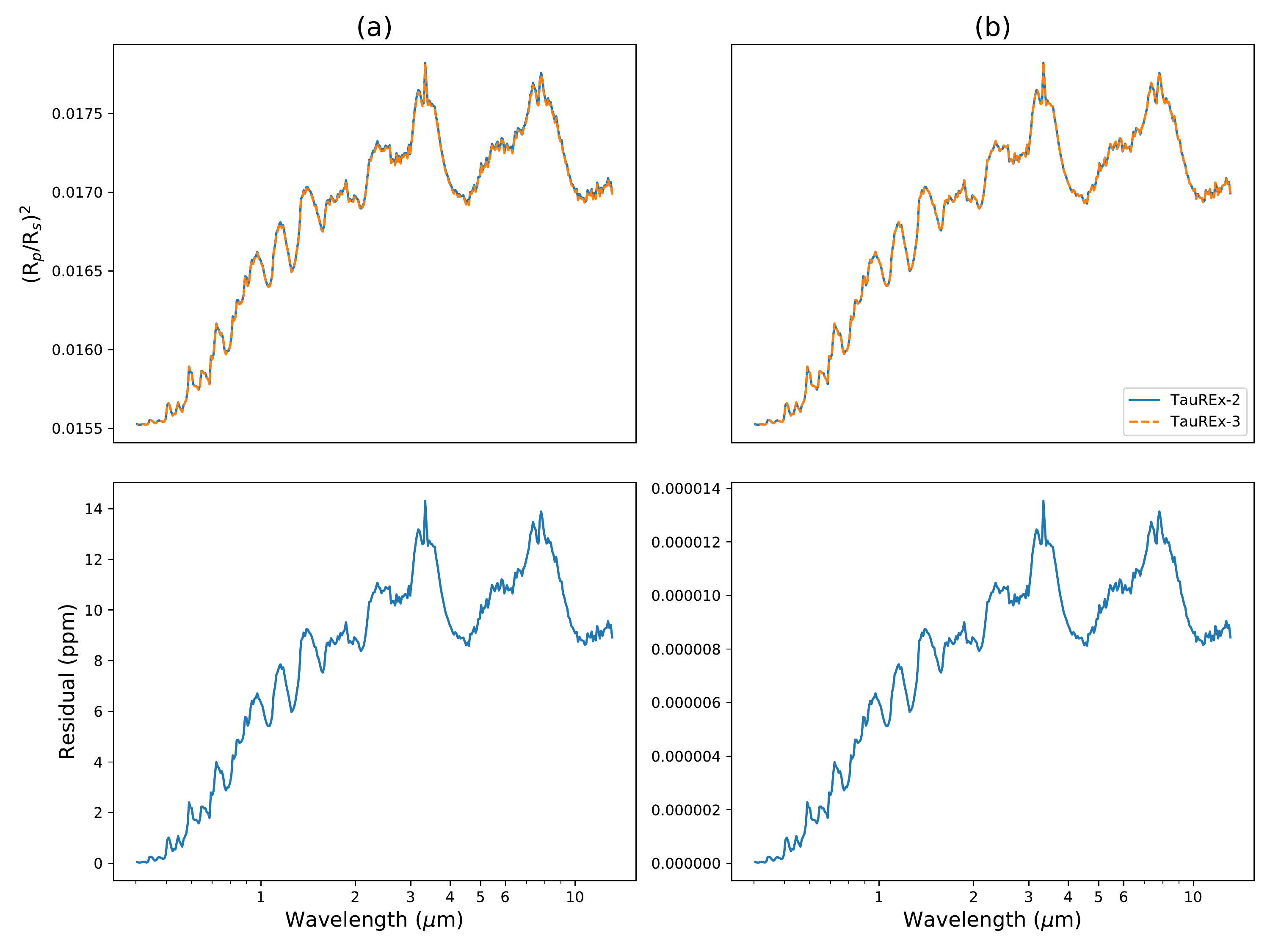}
\caption{Plots comparing transmission spectra computed at $R=10,000$ and binned to $R=100$ for clarity with their corresponding residuals in ppm. Both model the same planet and atmosphere and only include molecular absorption from H$_2$O and CH$_4$.
Plot (a) are both outputs given by their respective codes. Plot (b) compares \trex{3} with a modified \trex{2} that includes higher precision atomic and molecular mass values.}
\label{fig:transit_t2t3}
\end{figure*}

For the first benchmark, we verify the accuracy of spectra produced by \trex{3} when compared to \trex{2} for both, the transit and eclipse
cases. An arbitrary atmosphere is built which includes molecular absorption from H$_2$O and CH$_4$. Figure \ref{fig:transit_t2t3}a  compares both spectra and residuals of circa 15 ppm difference are observed. 

The majority of the residuals can be explained by the varying numerical precision of the atomic and molecular masses implemented in either code.
\trex{3} implements more precise values for the atomic and molecular masses
given in Table \ref{tab:masses}. This has the effect of compressing the atmosphere through the slightly heavier molecules, in turn
reducing the optical path length. We can observe this effect in figure \ref{fig:transit_t2t3}b when modifying the masses in \trex{2} to match those in \trex{3}. In this case, the residual differences become negligible ($\approx 10^{-5}$ ppm).

\begin{table*}[]
\centering
\begin{tabular}{lll}
\hline\hline
Atom/Molecule & \trex{2} mass (amu) & \trex{3} mass (amu) \\
\hline
H & 1 & 1.007940 \\
He & 4 & 4.002602 \\
H$_2$ & 2 & 2.01588 \\
H$_2$O & 18 & 18.01528 \\
CH$_4$ & 16 & 16.04276 \\

\hline
\end{tabular}
\caption{List of the atomic and molecular mass values used in Figure \ref{fig:transit_t2t3} between \trex{2} and \trex{3}}\label{tab:masses}
\end{table*}

\begin{figure*}[p]
\centering
    \includegraphics[width=0.98\textwidth]{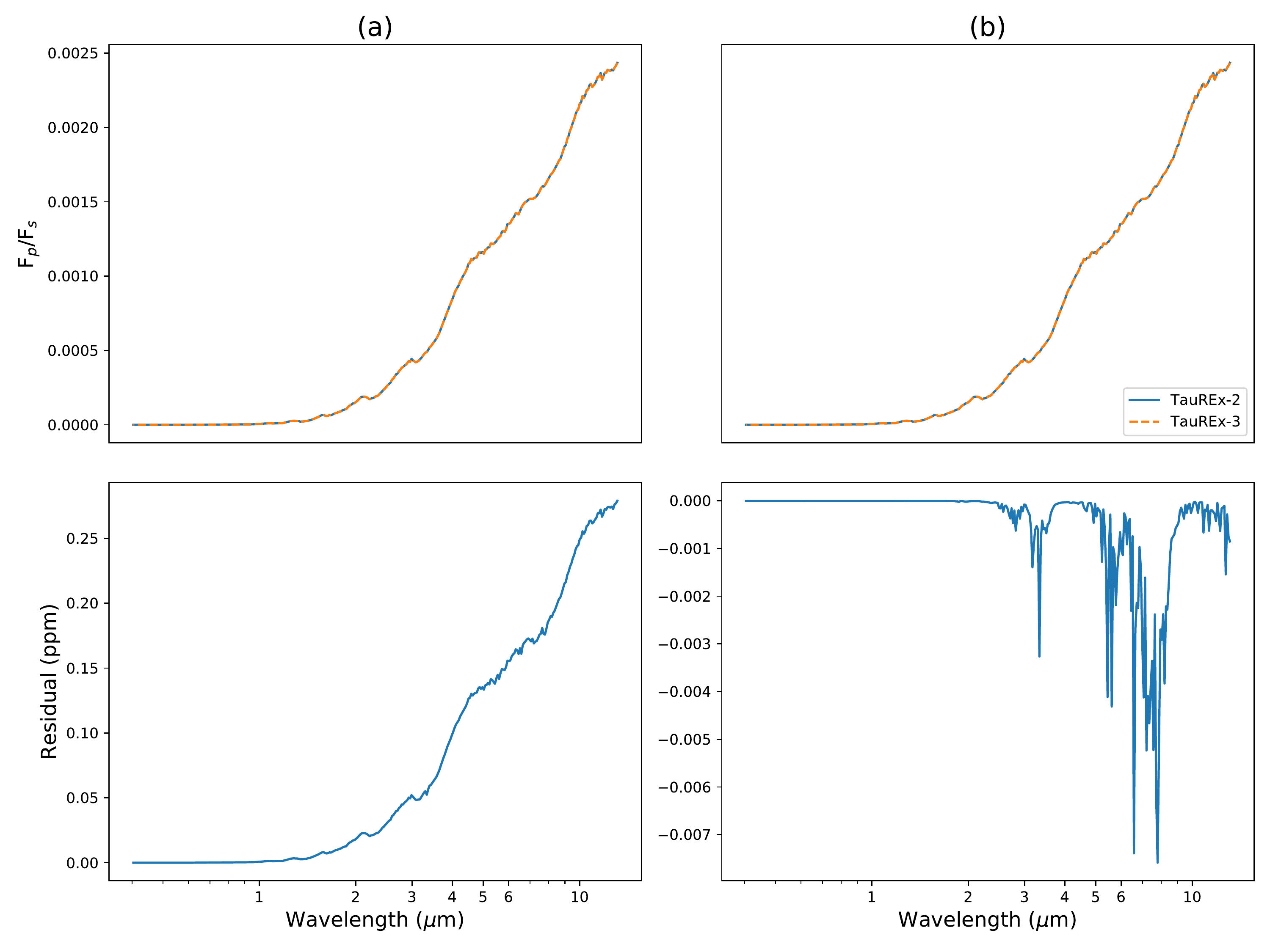}
\caption{Plots comparing 4-point quadrature emission spectra computed at $R=10,000$ and binned to $R=100$ for clarity with their corresponding residuals in ppm. Both model the same planet and atmosphere and only include molecular absorption from H$_2$O and CH$_4$.
Plot (a) compares \trex{3} with a modified \trex{2} that includes higher precision atomic and molecular mass values. Plot (b) compares \trex{3} with \trex{2} that has been further modified with higher precision
quadrature points}
\label{fig:eclipse_t2t3}
\end{figure*}

A similar exercise can be performed on the eclipse case. Figure \ref{fig:eclipse_t2t3}a) compares
both emission spectra from \trex{3} against the modified \trex{2} from Figure\,\ref{fig:transit_t2t3}b) as a baseline. The spectra match well with a residual of about 0.3\, ppm after removing the mass discrepancy between the codes. In the emission case, there are other sources of errors that arise from numerical precision. One example is the quadrature points, increasing the number of significant figures of the abscissa and weights in \trex{2}
to match those in \trex{3} further reduces the residuals by two orders of magnitude, as shown in Figure 
\ref{fig:eclipse_t2t3}b.

In summary, the more efficient path integral code has not degraded the quality of the output spectrum and the majority of differences between the two spectra stem from the higher precision constants used in \trex{3}.

The next set of benchmarks assess the performance scaling with the number of atmospheric layers for the full wavelength range. 
This step helps us to gauge the performance of the path-integral code with minimal influence from interpolation. 
Three methods were used; the first two were conducted using the previous version of \trex{2} using cross-sections and k-tables method, and the last is \trex{3} using cross-sections only. Each would build an atmospheric model with two active molecules with constant chemical profiles and an isothermal temperature profile. CIA with H$_2$-H$2$ and H$2$-He and Rayleigh scattering are included in the calculation.

Table \,\ref{tab:perf-fm-layer} demonstrates the significant performance upgrade from the previous version's cross-section code with 
an around 10$\times$ performance boost. For the 50 layer test, the interpolation time is the dominant computational bottleneck which gives the k-table method the advantage with its smaller opacity array. After this step, the path integral becomes the dominant computation time and \trex{3} matches the k-table method in performance at 100 layers. After this point, using \trex{3} with cross-sections 
is about 1.1 - 2$\times$ faster than \trex{2} with k-tables and around 54$\times$ faster
than the older cross-section code.

\begin{table}[]
\centering
\begin{tabular}{lrrrr}
\hline\hline
& \trex{2} & \trex{2}  &\trex{3} \\
Layers & xsec (s) & k-tables (s) & xsec (s) & \\
\hline
50     & 2.24         & 0.20     & 0.24    \\
100    & 8.60         & 0.79     & 0.62   \\
150    & 19.29        & 1.81     & 1.53        \\
200    & 35.53        & 3.04     & 2.29       \\
600    & 876.24       & 28.90    & 15.35  \\
\hline
\end{tabular}
\caption{A comparison of the forward model computation time between TauREx 2 using cross-sections, 
TauREx 2 using k-tables and TauREx 3 using cross-sections for the same atmospheric parameters but
increasing number of atmospheric layers.}\label{tab:perf-fm-layer}
\end{table}

Another essential comparison is how computation time scales with the number of molecules.
The same test was conducted but instead fixed at 100 atmospheric layers. Pseudo-molecules were generated by replicating the available cross-sections multiple times as different molecules. The results of table \ref{tab:perf-fm-molecules} again show
that k-tables perform best using a single molecule. With increasing number of molecules, \trex{3} performs significantly
better than both k-tables and the older cross-section code with a 2--8$\times$ and 8--100$\times$ performance gain respectively. 
\begin{table}[]
\centering
\begin{tabular}{lrrr}
\hline\hline
& \trex{2} & \trex{2}  &\trex{3} \\
Molecules & xsec (s) & k-tables (s) & xsec (s)  \\
\hline
1             & 7.23         & 0.45            & 0.61    \\
2             & 8.90          & 0.78           & 0.74       \\
4             & 12.42        & 1.49            & 0.92       \\
7             & 19.02           & 2.63         & 1.23        \\
15            & 263.56       & 8.21            & 2.34     \\
\hline
\end{tabular}
\caption{A comparison of the forward model computation time between TauREx 2 using cross-sections, 
TauREx 2 using k-tables and TauREx 3 using cross-sections for the same atmospheric parameters but
increasing number of molecules}\label{tab:perf-fm-molecules}
\end{table}




\subsection{Retrieval Benchmark: HD\,209458\,b}

For our retrieval benchmark, we will study HD\,209458\,b. Our first test will benchmark the current HST/WFC3 data, and the second will retrieve a simulated observation from ESA-Ariel mission \citep{Tinetti_ariel}.The aim is to assess both the consistency of the results and computational performance of \trex{3} against \trex{2}. We do not intend to perform a comprehensive study or reinterpretation of the available HD\,209458\,b data as this task would be beyond the scope of this paper.  We highlight that the previous version of TauREx\,2 was cross-compared with other retrieval codes in \cite{barstow_comparison}, showing the agreement between \trex{2}, the NEMESIS \citep{Irwin_nemesis} and Chimera \citep{Line_chimera} retrieval codes. 
For the optimizer, we use MultiNest \citep{Feroz_multinest,buchner_pymultinest} compiled with MPI. We utilize 1500 live points and an evidence tolerance of 0.5. The choice of hyperparameters ensures the adequate sampling of the retrieval's likelihood surface.
Each retrieval is done on a single node of the UCL cobweb cluster which has a 24-core Xeon E5-2697 v2 clocked at 2.70GHz. The timings are only for sampling and do not account for any startup time or post-processing.

For the first test, we compare the results of \trex{3} with the ones from \trex{2} in a real scenario for transmission and emission spectroscopy. We use the HST/WFC3 spectrum of HD\,209458\,b in \cite{Tsiaras_pop_study_trans} for our transmission scenario and the HST/WFC3 spectrum from \cite{Line_hd209em} for the emission case. For the latter case, we choose not to include the Spitzer points for our retrievals as combining instruments may lead to biases \citep{Yip_lightcurve}.

In our comparison retrieval, we attempt to constrain isocompositions for 5 molecules (H$_2$O, CH$_4$, CO, CO$_2$ and NH$_3$), using cross sections at a resolution of 10,000 given in Table \ref{tab:opacity-source}.

Along with the chemistry, we retrieve a temperature profile and the planet radius. In the transmission case, the temperature profile is isothermal and parameterized by a single parameter. We also retrieve the cloud-top pressure of a fully opaque cloud deck in the transmission case. 

In the emission case, we do not consider clouds as analysis by \cite{Line_hd209em} suggest their inclusion has a minimal impact on the dayside spectrum. We provide flexibility to the temperature using an N-point profile and retrieving three distinct values (the temperature at the surface (10 bar), at $1.5\times 10^{-1}$ bar and $2\times 10^{-3}$). We initially considered retrieving the pressure for these temperature points, but we found large degeneracies and decided to keep them fixed. The priors for the planet radius are kept narrow as they have some degree of degeneracy with temperature in the eclipse case due to the smaller wavelength coverage. In this scenario, preliminary knowledge (e.g the value obtained in the transit case) constrains the radius bounds as these observations are more sensitive to this parameter. By constraining the bounds, we can break the degeneracy and benchmark the retrieval of the N-point temperature profile against \trex{2}. 

For all these parameters, we use uniform priors which are listed in Table \ref{fig_priors}. 
\begin{table*}
\centering
$\begin{array}{l c c c }
\hline
\hline
\mbox{Retrieved Parameter} & \mbox{Transmission priors} & \mbox{Emission priors} & \mbox{Mode} \\
\hline
\mbox{H}_2\mbox{O} & \mbox{-12, -1} & \mbox{-12, -1} & \mbox{log} \\
\mbox{CH}_4 & \mbox{-12, -1} & \mbox{-12, -1} & \mbox{log} \\
\mbox{CO} & \mbox{-12, -1} & \mbox{-12, -1} & \mbox{log} \\
\mbox{CO}_2 & \mbox{-12, -1} & \mbox{-12, -1} & \mbox{log} \\
\mbox{NH}_3 & \mbox{-12, -1} & \mbox{-12, -1} & \mbox{log} \\
\mbox{T$_{isothermal}$ (K)} & \mbox{400, 2000} & \mbox{no} & \mbox{linear} \\
\mbox{T$_{surface}$ (K)} & \mbox{no} & \mbox{500, 2500} & \mbox{linear} \\
\mbox{T$_{point1}$ (K)} & \mbox{no} & \mbox{500, 2500} &  \mbox{linear} \\
\mbox{T$_{top}$ (K)} & \mbox{no} & \mbox{500, 2500} & \mbox{linear} \\
\mbox{radius (R$_J$)} & \mbox{1.2, 1.5} & \mbox{1.2, 1.5} & \mbox{linear} \\
\mbox{cloud pressure (bar)} & \mbox{1, -8} & \mbox{no} & \mbox{log} \\
\hline
\end{array}$
\caption{List of parameter retrieved in the transmission and emission retrieval along with their uniform bound priors and the retrieved mode.}
\label{fig_priors}
\end{table*}  

\begin{table*}[]
\centering
\begin{tabular}{lrrrr}
\hline\hline
Time (s) & \trex{2} (s) & \trex{3} (s) & No. samples & Speedup (x)  \\
\hline

Transit  & 6140      & 837       & 110,000  & 7.3 \\
Eclipse  & 3569    & 780       & 66,000   & 4.5 \\ 
\hline
\end{tabular}
\caption{A comparison of the retrieval model computation time between \trex{2} and \trex{3} using cross-sections for the same atmospheric priors}\label{tab:perf-ret}
\end{table*}

\begin{figure*}
\centering
    \includegraphics[width=0.49\textwidth]{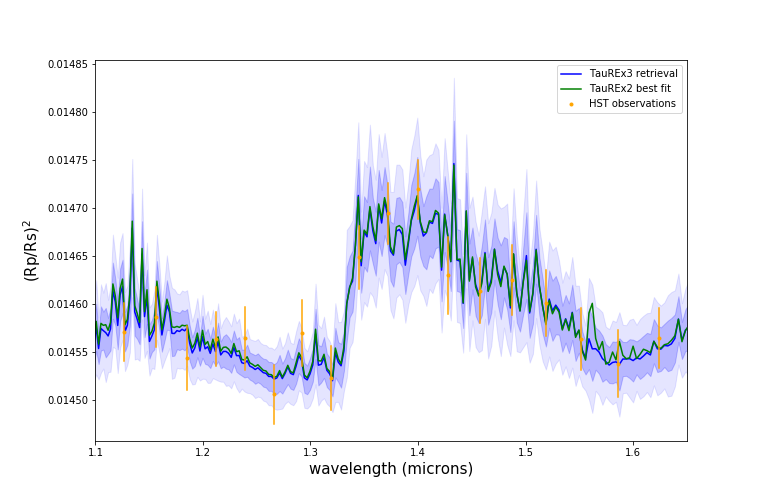}
    \includegraphics[width=0.49\textwidth]{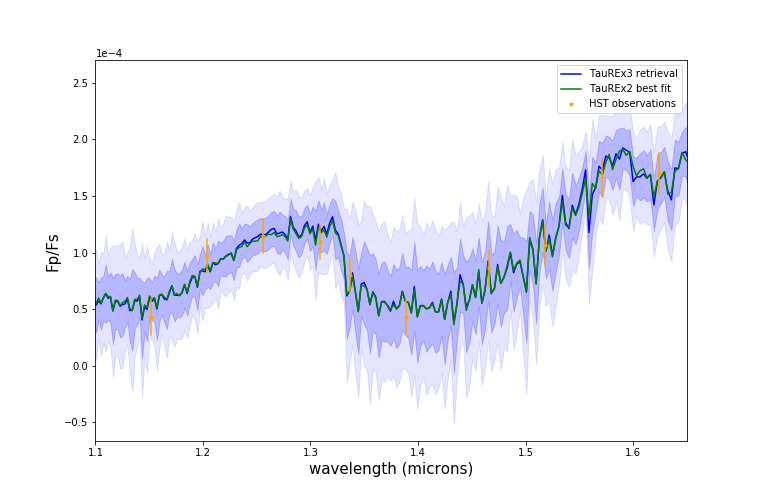}
\caption{Best fit spectra (denoted by solid lines) for the HD\,209458\,b retrievals with \trex{2} and \trex{3}. The 1\,$\sigma$ (dark shaded region) and 2\,$\sigma$ (light shaded region)  spectra are also plotted for the \trex{3} retrieval. Left: retrieval of the transmission spectrum from \cite{Tsiaras_pop_study_trans}. Right: retrieval of the emission spectrum from \cite{Line_hd209em}. The \trex{2} retrieval (green) and \trex{3} retrieval (blue) give the very similar spectra for both cases.}
\label{fig:spectra_ret}
\end{figure*}

The planet mass and star radius are fixed to the literature values, respectively 0.73 M$_J$ and  1.19 R$_\odot$ from \cite{Stassun_parameters}. On top of the five mentioned chemical species, we fill the rest of the atmosphere with hydrogen and helium at a ratio H$_2$/He = 0.17. Rayleigh scattering is calculated for all molecules \citep{cox_allen_rayleigh}, while we limit the collision-induced absorption to the couples H$_2$-H$_2$ and H$_2$-He. Finally, our model is computed in a grid of 100 layers with pressure ranging from 10 bar at the surface to $10^{-10}$ bar at the top of the considered atmosphere.
Figure\,\ref{fig:spectra_ret} shows the best fit spectra for \trex{2} and \trex{3} in the transmission and emission cases.

\begin{figure*}[p]
\centering
    \includegraphics[width=0.98\textwidth]{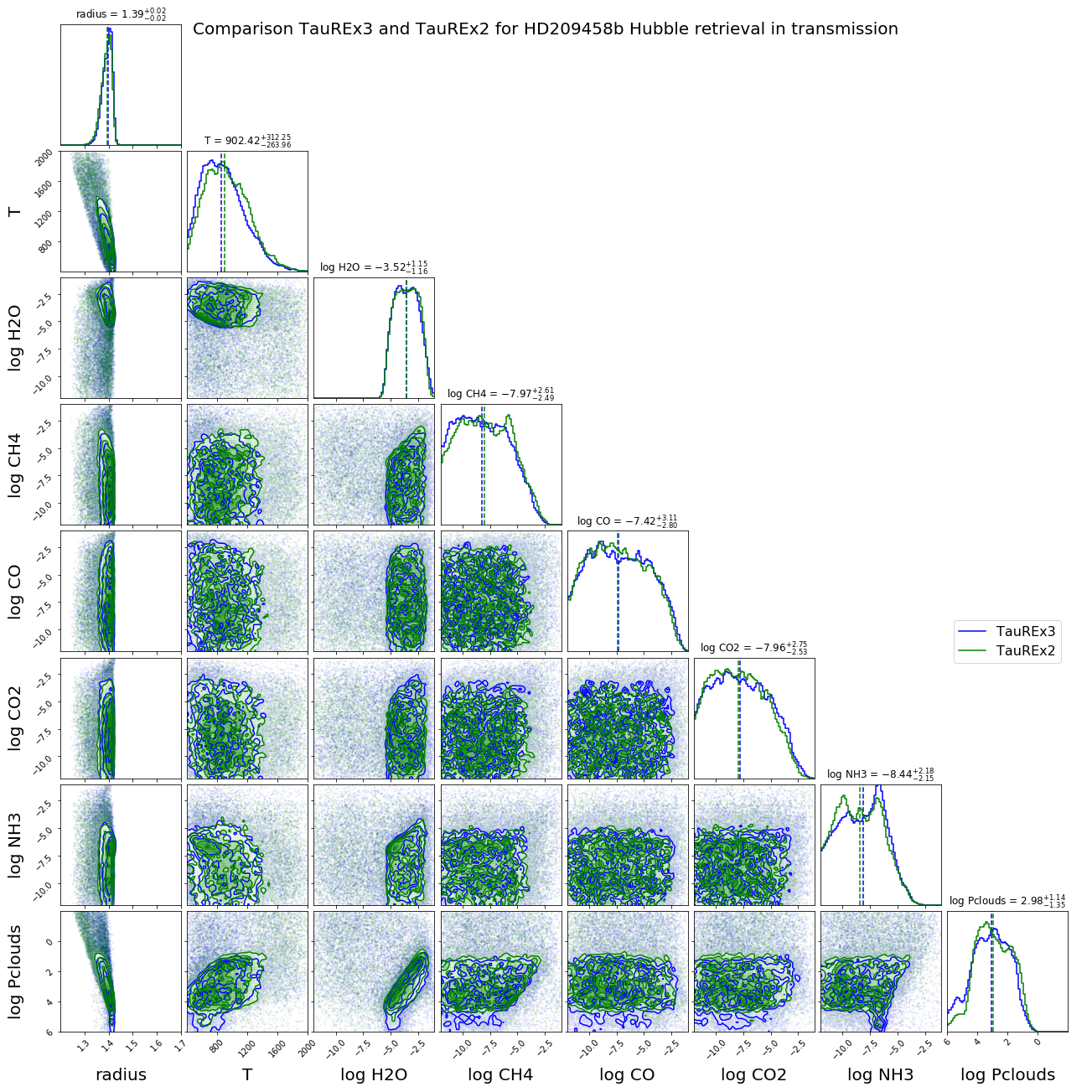}
\caption{Posterior distribution for the HD\,209458\,b retrievals of the transmission spectrum from \cite{Tsiaras_pop_study_trans} with \trex{2} and \trex{3}. Units of temperature are in Kelvin (K), pressures in Pascal (Pa), radii in Jupiter radius ($R_J$) and molecule abundance in volume mixing ratio (VMR). The dashed lines are the median values of the posterior. The values quoted above are the median from \trex{2} with 16\% and 84\% quantiles relative to it}
\label{fig:post_trans}
\end{figure*}

\begin{figure*}[p]
\centering
    \includegraphics[width=0.98\textwidth]{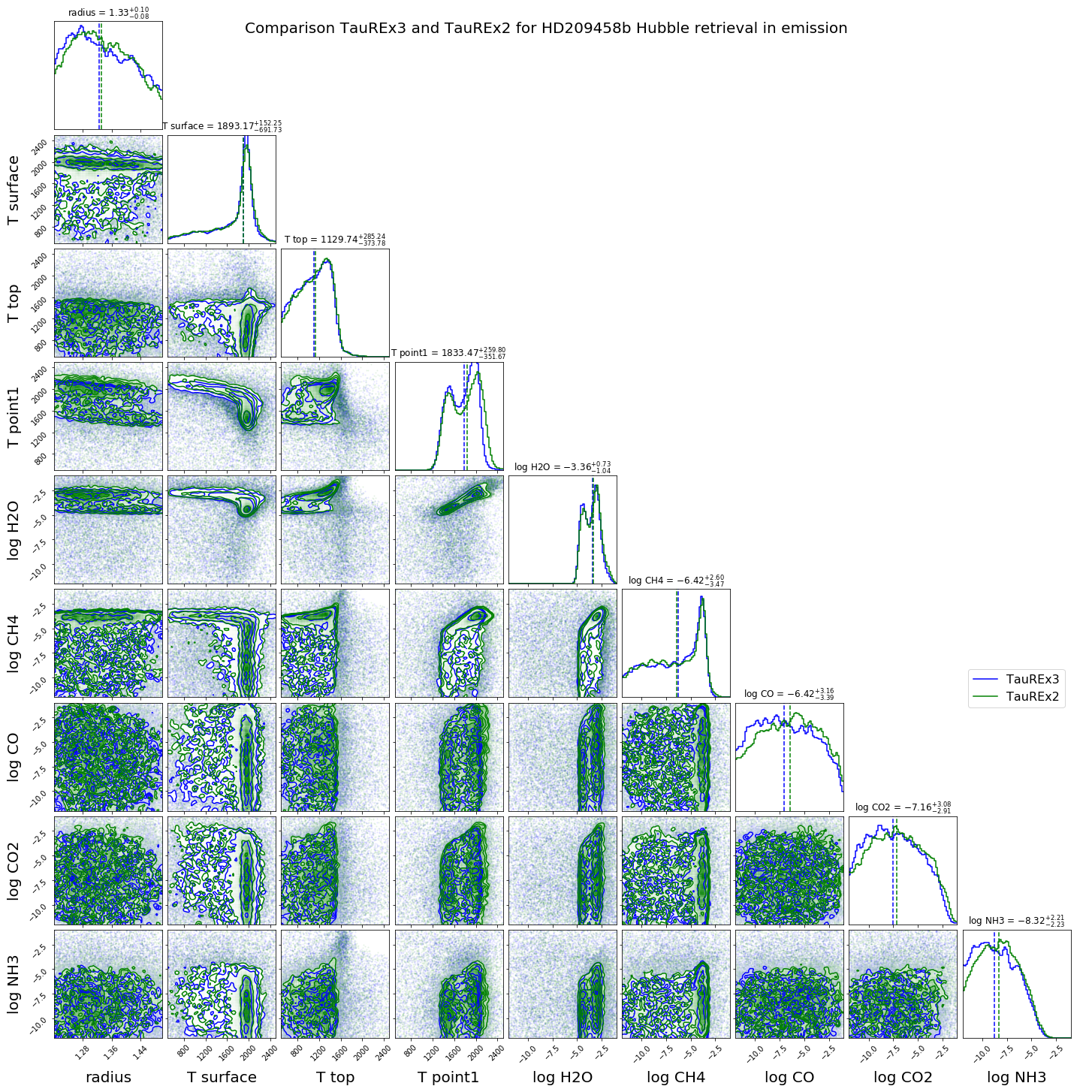}
\caption{Posterior distribution for the HD\,209458\,b retrievals of the emission spectrum from \cite{Line_hd209em} with \trex{2} and \trex{3}. Units of temperatures are in Kelvin (K), pressures in Pascal (Pa), radii in Jupiter radius ($R_J$) and molecule abundance in volume mixing ratio (VMR). The dashed lines are the median values of the posterior. The values quoted above are the median from \trex{2} with 16\% and 84\% quantiles relative to it}
\label{fig:post_em}
\end{figure*}

The posterior distributions for the transmission and emission cases are displayed in Figure \ref{fig:post_trans} and \ref{fig:post_em}. Retrieval times from Table \ref{tab:perf-ret} show a 4.5--7x speed up in sampling using TauREx 3 compared to TauREx 2.

From the spectra, posterior distributions and the retrieved parameters in Table \ref{tab:retrieval_params}, one can see that the results given by \trex{2} and \trex{3} are almost equivalent. The best fit spectra are within $1\,\sigma$ and the posterior distributions present the same shapes. Examining the parameters, we also see that all lie within 1\,$\sigma$ of each other's best fit values with variations arising only from the random sampling.


The two retrievals are also consistent with the main literature results. We retrieve H$_2$O in the dayside and terminator of the planet, respectively log(H$_2$O) = 3.36 and log(H$_2$O) = 3.52 in mixing ratios. In the terminator, we do not find significant evidence of additional molecules, but we retrieve a cloud pressure of about $10^{-2}$ bar. On the dayside, however, we find that the posterior distribution for CH$_4$ peaks around $10^{-4}$. This molecule seems to be degenerated with the temperature, which presents a double peak correlation. This result contrasts with both \cite{Line_hd209em} findings, in which they find evidence for CO. 

\begin{figure}[h]
\centering
    \includegraphics[width=0.5\textwidth]{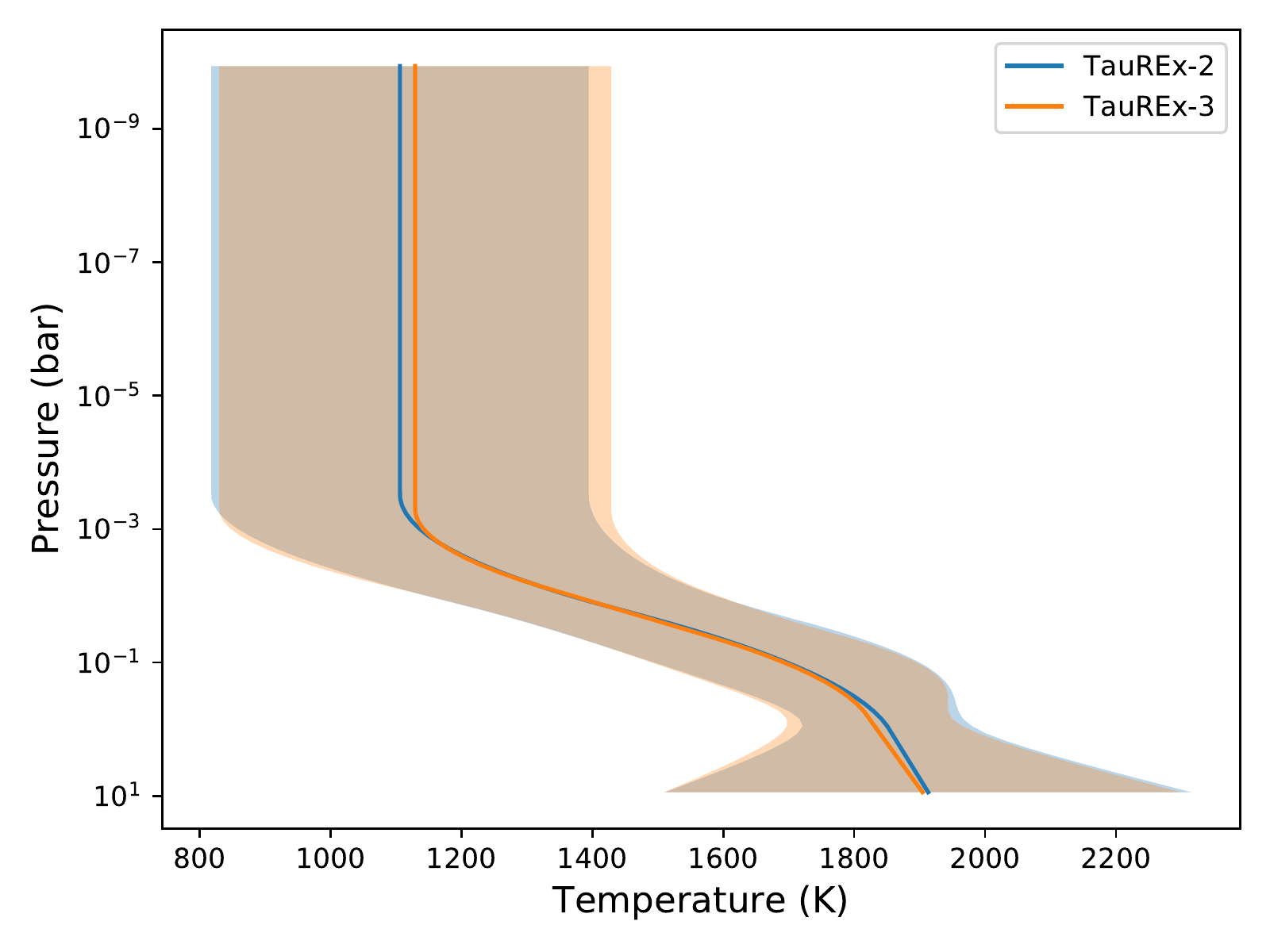}
\caption{Temperature profiles of day-side emission retrievals of HD\,209458\,b HST/WFC3 spectra \citep{Line_hd209em} by \trex{2} and \trex{3}. Solid lines denote the best fit values. Shaded ares denote the 1\,$\sigma$ span}
\label{fig:em_tp_prof}
\end{figure}

When conducting a retrieval using the 2-stream approximation temperature profile as described e.g. in \cite{Guillot2013}, we find similar constraint on H$_2$O but no constrain on either CO, CO$_2$ and CH$_4$. We speculate that the differences with results from \cite{Line_hd209em} come from including the Spitzer data in their analysis, which are more sensitive to carbon-bearing species.
Assessing the dayside temperature profile (Figure \ref{fig:em_tp_prof}) both retrievals give similar profiles and uncertainties with differences in temperature points of about 0.1 -- 60 K between each point.
\begin{table*}[p]
\centering
\begin{tabular}{lrrrr}
\hline\hline
 & Transmission &                & Emission & \\
Parameter & \trex{2} & \trex{3} & \trex{2} & \trex{3} \\
\hline
\\
log(H$_2$O)              & -3.52$^{+1.15}_{-1.16}$       & -3.57$^{+1.20}_{-1.21}$         & -3.35$^{+1.03}_{-0.72}$    & -3.47$^{+0.75}_{-0.98}$ \\
\\
log(CH$_4$)              & -7.97$^{+2.61}_{-2.49}$       & -7.87$^{+2.52}_{-2.59}$        & -6.42$^{+2.60}_{-2.47}$   & -6.41$^{+2.48}_{-3.64}$ \\
\\
log(CO)                  & -7.41$^{+3.10}_{-2.79}$       & -7.66$^{+3.22}_{-2.70}$        & -6.42$^{+3.16}_{-3.39}$   & -6.88$^{+3.54}_{-3.32}$ \\
\\
log(CO$_2$)              & -7.95$^{+2.74}_{-2.52}$       & -7.76$^{+2.73}_{-2.60}$        & -7.15$^{+3.07}_{-2.90}$  & -7.39$^{+3.03}_{-2.95}$ \\ 
\\
log(NH$_3$)              & -8.43$^{+2.18}_{-2.15}$       & -8.47$^{+2.19}_{-2.24}$        & -8.31$^{+2.21}_{-2.23}$  &  -8.46$^{+2.26}_{-2.16}$ \\
\\
T$_{isothermal}$ (K)     & 902.40$^{+312.22}_{-283.98}$  & 885.88$^{+352.88}_{-262.00}$   & -   & - \\
\\
T$_{surface}$ (K)        & -                             & -  & 1893.13$^{+152.27}_{-283.98}$  & 1904.30$^{+128.93}_{-606.58}$ \\
\\
T$_{point1}$ (K)         & -                             & -  & 1833.35$^{+259.90}_{-351.62}$   & 1773.91$^{+232.89}_{-304.51}$ \\
\\
T$_{top}$ (K)            & -                             & -  & 1129.62$^{+285.30}_{-373.69}$  & 1129.56$^{+286.15}_{-382.78}$ \\
\\
radius (R$_J$)           & 1.39$^{+0.01}_{-0.02}$        & 1.39$^{+0.01}_{-0.02}$         & 1.33$^{+0.09}_{-0.08}$   & 1.33$^{+0.09}_{-0.08}$ \\
\\
log(cloud pressure) (Pa) & 2.97$^{+1.13}_{-1.34}$        & 3.01$^{+1.28}_{-1.27}$        & -  & - \\ 
\\
$\mu$ (derived) & 2.29 $^{+0.09}_{-0.01}$  & 2.31$^{+0.07}_{-0.01}$ & 2.30$^{+0.13}_{-0.01}$ & 2.31$^{+0.09}_{-0.01}$ \\
\\
\hline \\
\end{tabular}
\caption{Retrieved parameters and their associated uncertainties for both emission and transmission of the HD\,209458\,b HST/WFC3 data for both \trex{2} and \trex{3}. $\mu$ is defined as the mean molecular weight of the atmosphere at the surface in atomic mass units. The Jupiter radius ($R_J$) is defined as $6.9911\times 10^{7}$~m }\label{tab:retrieval_params}
\end{table*}

For the second retrieval benchmark, we will use the \trex{2} best-fit values in Table \ref{tab:retrieval_params} and simulate a spectrum as seen from Ariel using ArielRad \citep{arielrad}. Our wavelength coverage ranges from 0.5--7.8\,\micron, effectively increasing the number of wavelength bins six-fold compared to the HST/WFC3 case. The increase in information will impact the computational cost in calculating the forward model at each iteration and the number of samples required to achieve adequate convergence during retrievals.

\begin{table*}[p]
\centering
\begin{tabular}{lrrrr}
\hline\hline
Time (s) & \trex{2} (s) & \trex{3} (s) & No. samples & Speedup (x)  \\
\hline

Transit  & 42,145      & 6,885     & 180,000  & 6.2 \\
Eclipse  & 72,607   & 10,559   & 150,000   & 6.87 \\ 
\hline
\end{tabular}
\caption{A comparison of the retrieval model computation time between TauREx 2 and TauREx 3 using cross-sections for the same atmospheric priors computed on Ariel simulated spectra} \label{tab:perf-ret-ariel}
\end{table*}

Comparing the retrievals in Table \, \ref{tab:perf-ret-ariel}, we indeed see a significant increase in the number of samples
required and the time taken to complete. Highlighted is the significantly reduced retrieval time with \trex{3} for both transit
and eclipse times requiring only 2 and 3 hours respectively compared to \trex{2} retrievals taking 11 and 20 hours respectively. Not including the fact that the original code takes almost 30 minutes to
set up before starting a retrieval compared to seconds for the current version.
Scaling up to Ariel resolution introduces a 6$\times$ increase in single sampling times for \trex{3} which is in-line with the increase in resolution. In contrast to the 20$\times$ increase in runtime for \trex{2}. 

\begin{figure}
\centering
    \includegraphics[width=0.8\columnwidth]{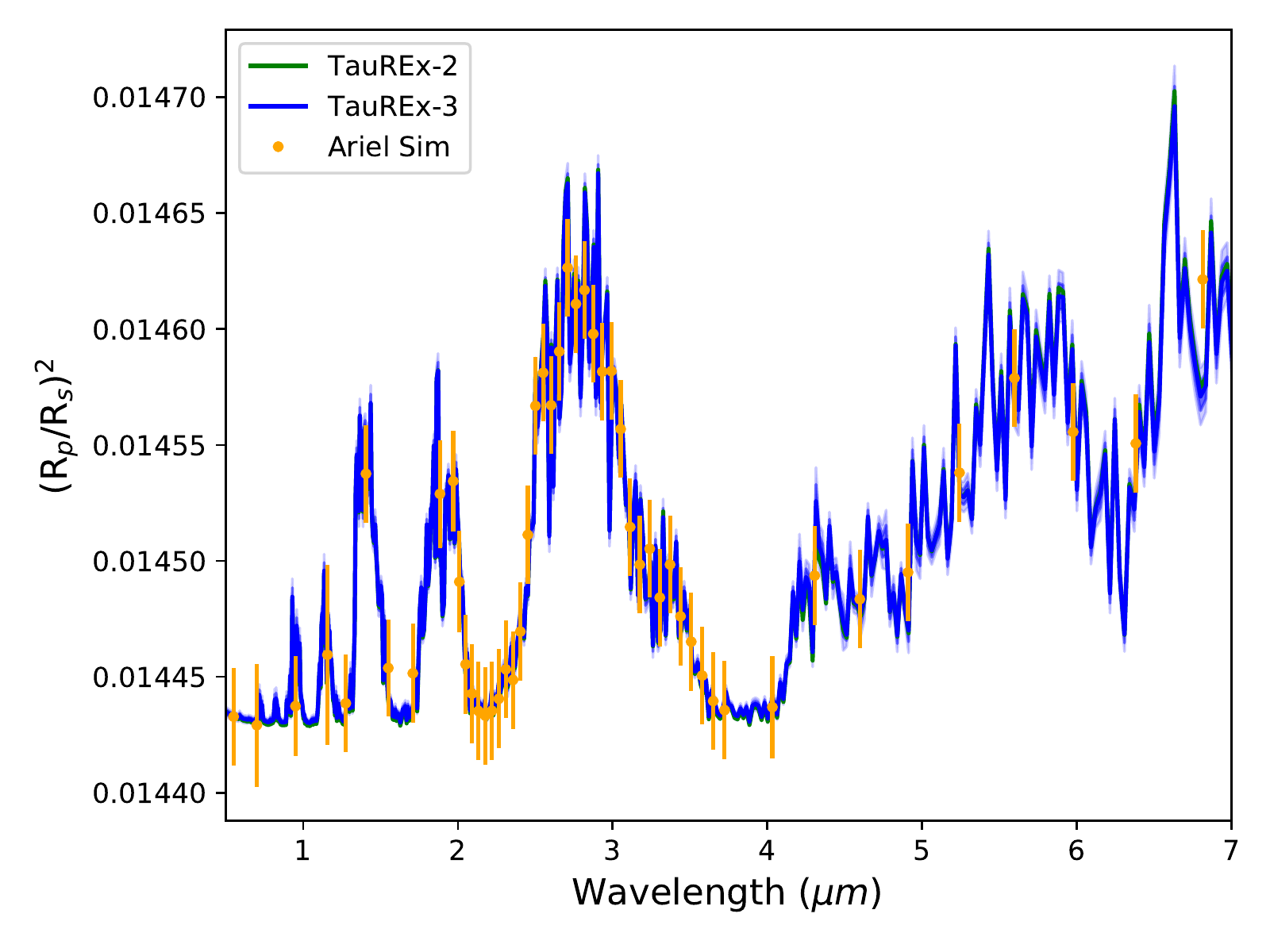}
    \includegraphics[width=0.8\columnwidth]{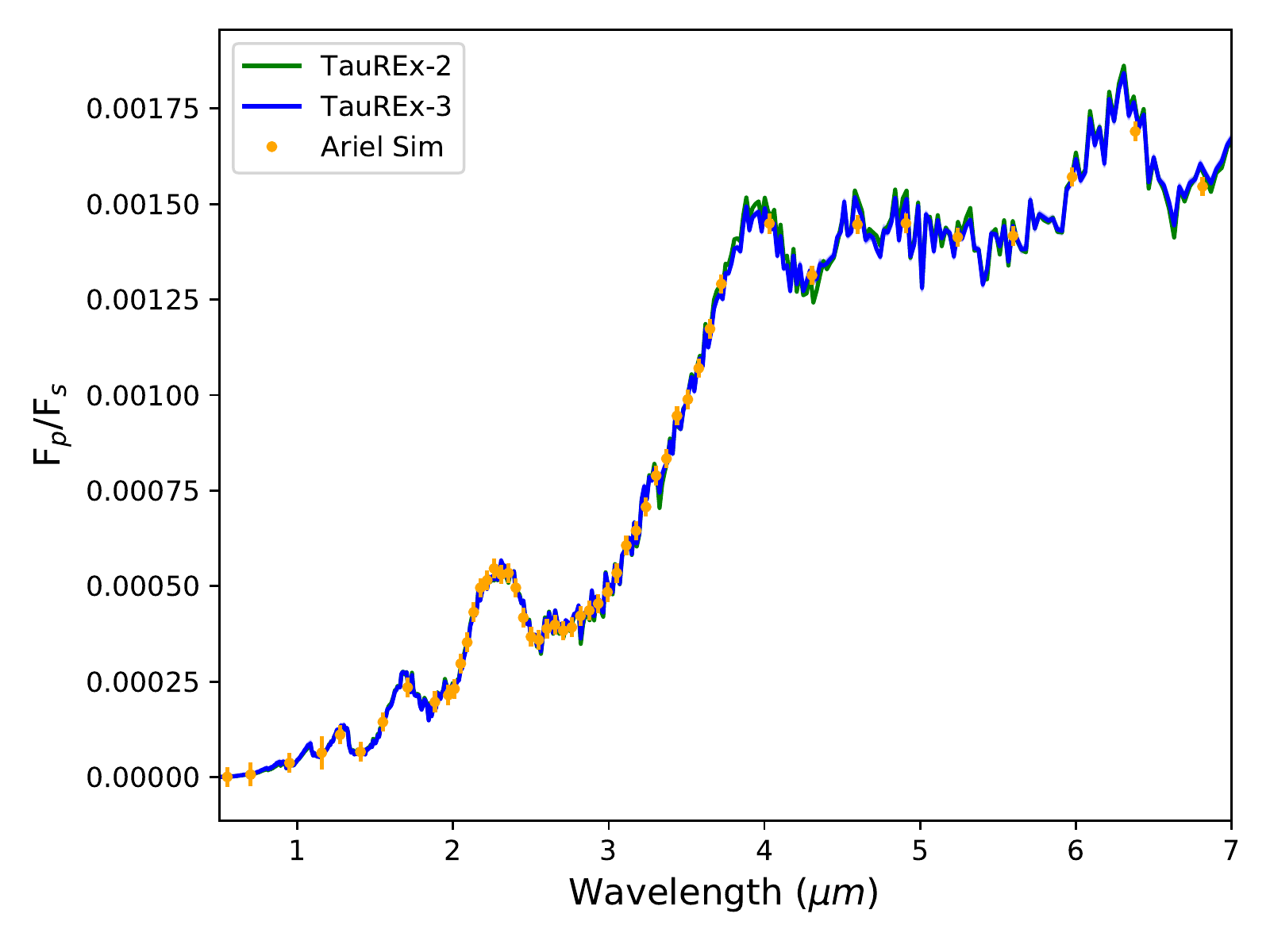}
\caption{Best fit spectra for the HD\,209458\,b retrievals with \trex{2} and \trex{3}. The 1\,$\sigma$ and 2\,$\sigma$ spectra are also plotted for the \trex{3} retrieval. Top: retrieval of the transmission spectrum from simulated Ariel spectrum. 
Bottom: retrieval of the emission spectrum from simulated Ariel spectrum. The simulated spectrum for each plot are generated from the best fit values of \trex{2} from Table \ref{tab:retrieval_params} with noise simulated by ArielRad\citep{arielrad}}
\label{fig:spectra_ret_ariel}
\end{figure}

\begin{figure*}[p]
\centering
    \includegraphics[width=0.98\textwidth]{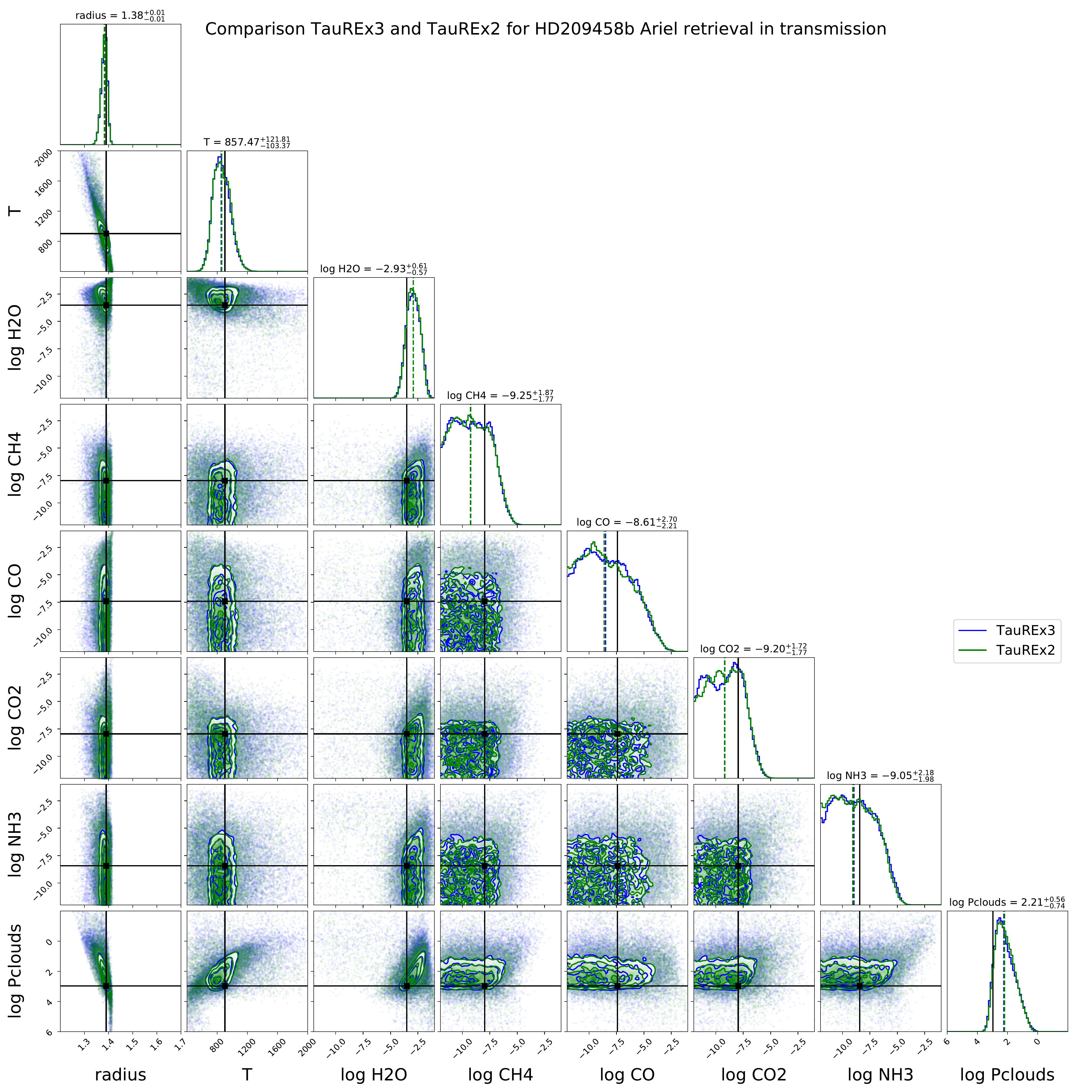}
\caption{Posterior distribution for the HD\,209458\,b retrievals of the transmission spectrum on HD209458 b simulated Ariel observation with \trex{2} and \trex{3}. The simulated spectrum uses the best fit values of \trex{2} from Table~\ref{tab:retrieval_params} (solid black lines). The dashed lines are the median values of the posterior. The values quoted above are the median from \trex{2} with 16\% and 84\% quantiles relative to it}
\label{fig:post_trans_ariel}
\end{figure*}

\begin{figure*}[p]
 \centering
     \includegraphics[width=0.98\textwidth]{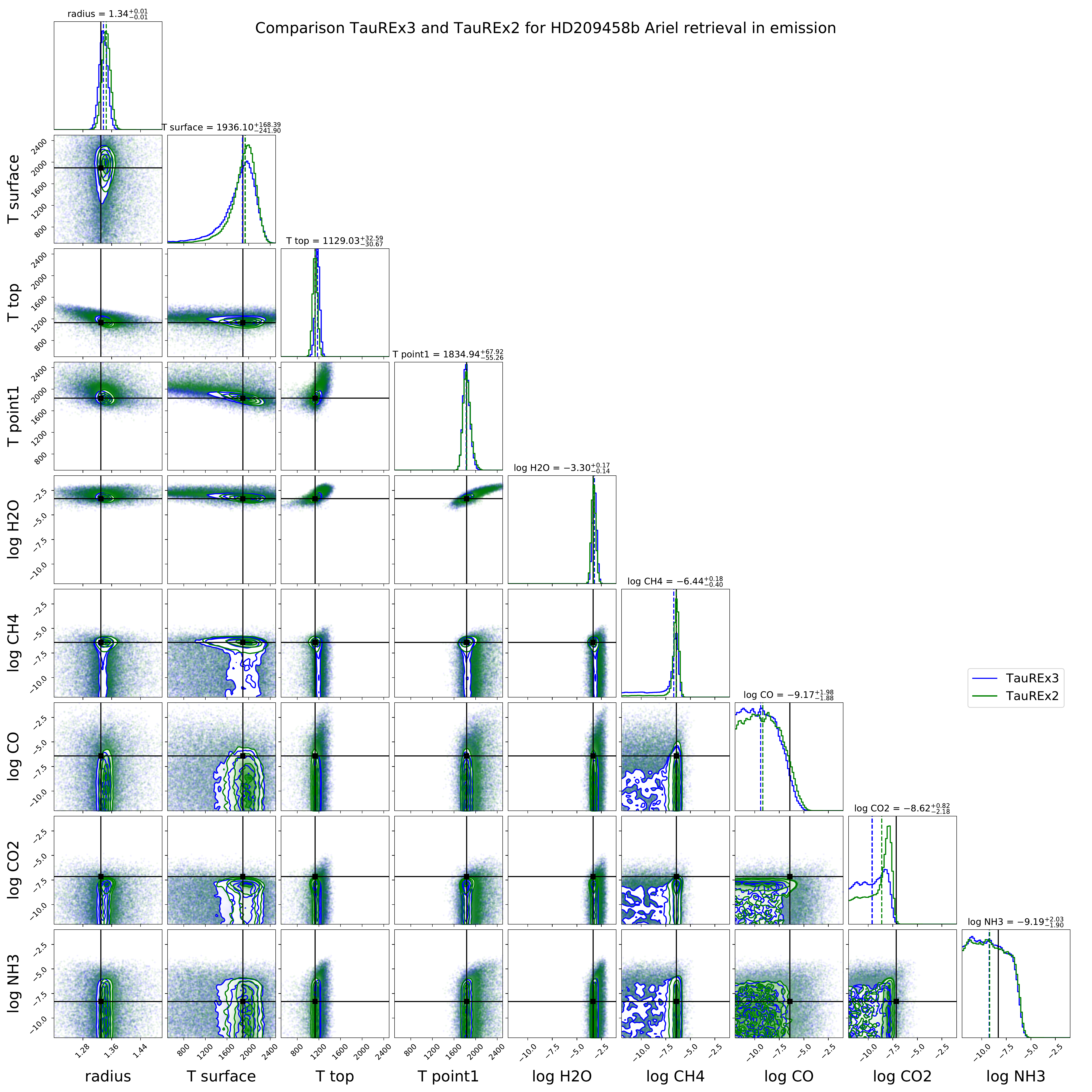}
 \caption{Posterior distribution for the HD\,209458\,b retrievals of the emission spectrum on HD209458 b simulated Ariel observation with \trex{2} and \trex{3}. The simulated spectrum uses the best fit values of \trex{2} from Table \ref{tab:retrieval_params} (solid black lines). The dashed lines are the median values of the posterior. The values quoted above are the median from \trex{2} with 16\% and 84\% quantiles relative to the median}
 \label{fig:post_em_ariel}
 \end{figure*}

Examining the spectra in Figure\,\ref{fig:spectra_ret_ariel} we see that both are essentially identical. This result is expected since the greater resolution and smaller
SNR increases the information available to the retrieval and greatly helps to break degeneracies. This behaviour is evident in the posterior distributions
given in Figures\,\ref{fig:post_trans_ariel} and \ref{fig:post_em_ariel} where most parameters are well constrained and within 1$\sigma$ of the truth values. The emission posteriors for the radius, T$_{point 1}$ and T$_{top}$ are now well defined. As an aside, this highlights how dedicated missions such as Ariel present a significant improvement to our ability to resolve spectral features in exoplanets.  Some of the truth values, namely $P_{clouds}$, log CO and log CO$_2$, lie outside the 1-$\sigma$ retrieved values. This result is expected as the retrieval traces the information content and degeneracies. For example, CO does not produce visible features in our simulated spectrum, meaning the retrieval only recovers an upper limit. The calculated mean and 1-$\sigma$ for these parameters become prior dependant. We refer the interested reader to literature discussing retrieval correlation in further detail 
\citep[e.g.][]{Feng_2016,Rocchetto_jwst, changeat2019complex, alfnoor,changeat_mass}.

In summary, benchmarking shows results of \trex{3} to be consistent with \trex{2} (which, in turn, was benchmarked against both NEMESIS and CHIMERA \cite{barstow_comparison}), while giving a 6-fold improvement in retrieval runtime. This result demonstrates that retrievals of higher resolution spectra from missions such as JWST and Ariel are feasible within a couple of hours.


\section{Future Work}
\label{sec:future}
The flexibility afforded in \trex{3} will allow for a wide range of novel applications to be developed on top of its core functionality. 
In future publications, currently in preparation, we will present new applications using this library. The list includes: a better treatment of scattering processes with a two-stream \citep{twostream} and multi-stream approximations \citep{disort} for the emission model; a new forward model generation and retrieval pipeline for large scale studies of planetary populations for next-gen telescopes \citep{alfnoor}; applications to solar system bodies with solar occultation \citep{tgo}, nadir models and integration of the Mars Climate Database (MCD) chemistry model \citep{mcd};
a new phase curve forward model for exoplanetary applications.

In general, the design of \trex{3} aims to reduce significantly  the time and effort for other groups to include their own chemistry/clouds/forward/temperature schemes. The framework is fully open source, under a BSD licence, and we hope to provide a community wide tool for retrieval code development in the future.


\section{Summary}

In this publication, we present our new retrieval framework \trex. 
The code can act as a library providing ready to use functions for atmospheric
modelling. These components can combine into a full pipeline for atmospheric
retrievals. \trex{3} is flexible, seamlessly utilizing new codes defined by
the user. We demonstrated its dynamic nature, responding to changes in the forward model and generating new and appropriate fitting parameters for retrievals. \trex{3} is designed to adapt to the rapid development of atmospheric theory
and include the cutting edge with minimal effort. It allows for the rapid prototyping of
new methods and retrieval regimes. This includes more complex retrievals such as observation geometry and model hyper-parameters.
Our benchmarks demonstrate the significant improvement in performance at high resolution ($R=10,000$) with
improvements in performance, reaching 10$\times$ the previous version at large wavelength ranges. Retrieval times are significantly reduced with simulated Ariel retrievals completing in a couple of hours.
Our benchmarks also demonstrate its robustness and show that the results of \trex{3} match precisely with the previous version.
The code is open-source, licensed under a BSD license and available on Github \footnote{\url{http://github.com/ucl-exoplanets/TauREx3_public}} and the Python Package Index (PyPi) \footnote{\url{https://pypi.org/project/taurex/}}. \\

\noindent\textbf{Acknowledgements}

This project has received funding from the European Research Council (ERC) under the European Union's Horizon 2020 research and innovation programme (grant agreement No 758892, ExoAI) and the European Union's Horizon 2020 COMPET programme (grant agreement No 776403, ExoplANETS A). Furthermore, we acknowledge funding by the UK Space Agency and Science and Technology Funding Council (STFC) grants: ST/K502406/1, ST/P000282/1, ST/P002153/1, ST/S002634/1, ST/T001836/1 and ST/V003380/1.

\bibliographystyle{aasjournal}  
\bibliography{references}


\end{document}